\documentclass[apj,twocolumn]{emulateapj}
\usepackage{natbib}
\usepackage{float}
\usepackage{graphics,rotating,multirow,bm,color,amsmath,amssymb,hyperref}
\usepackage{epsfig}

\hypersetup{
    colorlinks=true,
    linkcolor=green,
    citecolor=blue,
}

\begin{document}

\title{$N$-Body Simulations for Extended Quintessence Models}
\author{Baojiu~Li$^{1,2}$, David~F.~Mota$^3$ and  John~D.~Barrow$^2$}
\email[Email address: ]{b.li@damtp.cam.ac.uk}
\affiliation{$^1$DAMTP, Centre for Mathematical Sciences, University of Cambridge,
Wilberforce Road, Cambridge CB3 0WA, UK}
\affiliation{$^2$Kavli Institute for Cosmology Cambridge, Madingley Road, Cambridge CB3 0HA,
UK}
\email[Email address: ]{d.f.mota@astro.uio.no}
\affiliation{$^3$Institute of Theoretical Astrophysics, University of
Oslo, 0315 Oslo, Norway}
\email[Email address: ]{j.d.barrow@damtp.cam.ac.uk}

\begin{abstract}
We introduce the $N$-body simulation technique to follow
structure formation in linear and nonlinear regimes for the extended
quintessence models (scalar-tensor theories in which the scalar field has a
self-interaction potential and behaves as dark energy), and apply it
to a class of models specified by an inverse power-law potential
and a non-minimal coupling. Our full solution of the scalar field
perturbation confirms that, when the potential is not too
nonlinear, the effects of the scalar field could be accurately
approximated as a modification of background expansion rate plus a
rescaling of the effective gravitational constant relevant for
structure growth. For the models we consider, these have opposite
effects, leading to a weak net effect in the linear perturbation
regime. However, on the nonlinear scales the modified expansion
rate dominates and could produce interesting signatures in the
matter power spectrum and mass function, which might be used to
improve the constraints on the models from cosmological data. We
show that the density profiles of the dark matter halos are well
described by the Navarro-Frenk-White formula, although the scalar
field could change the concentration. We also derive an analytic
formula for the scalar field perturbation inside halos assuming
NFW density profile and sphericity, which agrees well with
numerical results if the parameter is appropriately tuned. The
results suggest that for the models considered, the spatial
variation of the scalar field (and thus the locally measured
gravitational constant) is very weak, and so local experiments
could see the background variation of gravitational constant.
\end{abstract}

\maketitle

\section{Introduction}

\label{sect:intro}

The nature of the dark energy \citep{Copeland:2006} is one of the
most difficult challenges facing physicists and cosmologists now.
Although a cosmological constant (plus cold dark matter, to provide
the concordance $\Lambda$CDM paradigm) could be a solution -- and is
indeed consistent with virtually all current observations, it
suffers from theoretical difficulties such as why its value must
be so small yet nonzero, and why it becomes dominant only at the
low redshift. In all the alternative proposals to tackle this
problem, a quintessence scalar field \citep{Zlatev:1999, Wang:2000}
is perhaps the most popular one (although a new proposal by \cite{Barrow:2010} provides a
completely new type of explanation that does not require new scalar fields). In such models the scalar field
$\varphi$ is slowly rolling down its potential, its energy density is
dominated by the potential energy and almost remaining constant
provided that the potential is flat enough. The flatness of the
potential, however, means that the mass of the scalar field is in
general very light and as a result the scalar field almost does
not cluster so that its effects in cosmology are mainly on the
(modified) background expansion rate.

One reason for the wide interest in quintessence models is that
scalar fields appear in abundance in high-energy physics theories, in which they are often coupled to the curvature invariants or even other matter species, leading to
the so-called extended quintessence \citep{Perrotta:1999, Baccigalupi:2000, Baccigalupi:2000b} 
and coupled quintessence \citep{Amendola:2000, Amendola:2004, Jesus:2008} models respectively. The former is just a special class of a
scalar-tensor theory \citep{scalar_tensor, Riazuelo:2002},
with the scalar field being the dark energy. These two classes of
generalised quintessence models have been studied in detail in
the linear regime in the literature
\citep{Bean:2001a, Bean:2001b, Mangano:2003, Clifton:2004, Nunes:2004, Pettorino:2005, Koivisto:2005, Brookfield:2005, Koivisto:2006, Mota:2006, Mota:2006c, Lee:2006, Mota:2007b, Bean:2008a, Bean:2008b, Boehmer:2008, Boehmer:2010}.

In recent years, studies of the cosmological behaviour of the coupled
quintessence model in the nonlinear regime have also been made, either via semi-analytical methods
\citep{Manera:2005, Mota:2004a, Mota:2008, Shaw:2007, Mota:2008b, Mota:2007d, Saracco:2010, Wintergerst:2010},
or using $N$-body simulation techniques
\citep{Maccio:2004, Nusser:2005, Kesden:2006, Kesden:2006b, Springel:2007, Farrar:2007, Baldi:2010, Hellwing:2009,
Keselman:2009, Keselman:2010, Hellwing:2010, Baldi:2010a, Baldi:2010b}. In these studies the effect of the scalar field is generally
approximated by a Yukawa-type 'fifth force' or by a
rescaling of the gravitational constant or the particle mass, without
solving the scalar field equation explicitly. Very recently, \cite{Li:2009sy, Li:2010, Zhao:2010, Li:2010b} gave a new treatment and
obtained an explicit solution to the scalar field perturbation on a
spatial grid. The new results confirmed that the approximations adopted in
the old literature were good for the models considered there (where
the scalar potential was not very nonlinear), but for highly nonlinear
potentials they broke down.

For the extended quintessence (more generally scalar-tensor) models,
investigations using $N$-body simulations are rarer. The work of \cite{Pettorino:2008}, for example, outlined a recipe which uses certain
approximation, such as a rescaling of gravitational constant, and
does not solve the scalar field equation of motion explicitly. In \cite{Rodriguez-Meza:2007, Rodriguez-Meza:2008a, Rodriguez-Meza:2008b}, the
authors approximated the effect of scalar field coupling as a Yukawa force.
However, none of these previous works tries to solve the scalar field on a
mesh directly, and this is what we want to do in this work.

The aims of this work are threefold. Firstly, we want to
develop the formulae and methods that are needed to solve the scalar
field explicitly, which could serve as the basis for future
work, and to find the regime of validity of our method. Secondly, we want
to understand whether the approximations adopted in the previous
studies are good or not; given the severe limits in the computing power; if those approximations do work well, then one does not need to resort to
an exact scalar field solver, which is considerably more economical.
Finally, we want to study structure formation in the nonlinear
regime for some specific models, and investigate both the scalar
field effects on the clustering of matter and the spatial variation of the
gravitational constant (which is common to scalar-tensor
theories).

The organisation of this paper is as follows: In Sect.~\ref{sect:eqn} we
list the basic equations which are needed in $N$-body simulations and give
their respective non-relativistic limits. To prevent the main text from
expanding too much, some useful expressions are listed in Appendix~\ref{appen:expression}, and the discrete versions of the resulted equations are
discussed and summarised in Appendix~\ref{appen:discret}. In Sect.~\ref{sect:simu} we briefly describe the numerical code we are using
(relegating further details to \cite{Knebe:2001, Li:2010}), and the
physical parameters of our simulations. We also present some results
regarding the background cosmology and linear perturbation evolution
in our models, which could be helpful in the understanding of the $N$-body simulation results (our algorithm for the background cosmology
is summarized in Appendix~\ref{appen:bkgd}). Sect.~\ref{sect:results}
contains the $N$-body simulation results, including key structure formation
observables such as nonlinear matter power spectrum, mass function and dark
matter halo profile, as well as the spatial variation of the scalar
field. It also includes several checks of the approximations made in the
literature. We finally summarise and conclude in Sect.~\ref{sect:con}.

We use the unit $c=1$ unless explicitly restoring $c$ in the
equations. The metric convention is $(+,-,-,-,)$. Indices $a,b,c,\cdots $ run $0,1,2,3$ while $i,j,k,\cdots $ run $1,2,3$.

\section{The Equations}

This section presents the equations that will be used in the
$N$-body simulations, the model parameterisation and
discretisation procedure for the equations.

\label{sect:eqn}

\subsection{The Basic Equations}

\label{subsect:basiceqn}

We consider a general Lagrangian density for scalar-tensor
theories
\begin{eqnarray}\label{eq:Lagrangian}
\mathcal{L} &=& \frac{1}{2\kappa_{\ast}}\left[1+f(\varphi)\right]R
- \frac{1}{2}\nabla^{a}\varphi\nabla_{a}\varphi + V(\varphi) -
\mathcal{L}_{f},
\end{eqnarray}
in which $\kappa_{\ast}=8\pi G_{\ast}$ where $G_{\ast}$ is the
(bare) gravitational constant, $R$ is the Ricci scalar,
$f(\varphi)$ is the coupling function between the scalar field
$\varphi$ and curvature, $V(\varphi)$ the potential for $\varphi$
and $\mathcal{L}_{f}$ the Lagrangian density for fluid matter
(baryons, photons, neutrinos and cold dark matter). Note that
$G_{\ast}$ is a fundamental constant of the theory.

Varying the associated action with respect to metric $g_{ab}$ yields the energy-momentum tensor of the theory
(note the tilde, which is used to distinguish it from the $T_{ab}$
defined below):
\begin{eqnarray}\label{eq:emt}
\tilde{T}_{ab} &=& T^{f}_{ab} + \nabla_{a}\nabla_{b}\varphi -
\frac{1}{2}g_{ab}\nabla^{c}\varphi\nabla_{c}\varphi +
g_{ab}V(\varphi)\nonumber\\
&&-\frac{1}{\kappa_{\ast}}\left[f(\varphi)G_{ab} +
\left(g_{ab}\nabla^{c}\nabla_{c}-\nabla_{a}\nabla_{b}\right)f(\varphi)\right]
\end{eqnarray}
where $G_{ab}=R_{ab}-\frac{1}{2}g_{ab}R$ is the Einstein tensor,
and $T^{f}_{ab}$ is the energy-momentum tensor for matter
(including baryons, dark matter, neutrinos and photons, which we
collectively refer to as 'fluid matter', although in $N$-body
simulations we use discrete particles rather than a fluid).

As usual, we can rearrange the Einstein equation as
\begin{eqnarray}
G_{ab} &=& \kappa_{\ast}\tilde{T}_{ab}
\end{eqnarray}
so that it now looks like
\begin{eqnarray}\label{eq:Einstein}
G_{ab} &=& \frac{\kappa_{\ast}}{1+f}T^{f}_{ab} -
\frac{1}{1+f}\left(g_{ab}\nabla^{c}\nabla_{c}-\nabla_{a}\nabla_{b}\right)f\nonumber\\
&&+\frac{\kappa_{\ast}}{1+f}\left[\nabla_{a}\varphi\nabla_{b}\varphi
- \frac{1}{2}g_{ab}\left(\nabla\varphi\right)^2 +
g_{ab}V\right]\nonumber\\
&\equiv& \kappa_{\ast}T_{ab}.
\end{eqnarray}
Note the difference between $T^{f}_{ab}$ and $T_{ab}$; throughout
this paper, we will use a superscript $^f$ for
normal fluid matter, and quantities without a superscript $^f$
always mean the total effective ones [the final line of
Eq.~(\ref{eq:Einstein})]. It is sometimes useful to define an
effective Newton constant
$\kappa_{eff}\equiv\kappa_{\ast}/(1+f)$. Neither
$\kappa_\ast$ nor $\kappa_{eff}$ is the gravitational constant
measured in a Cavendish-type experiment, which we denote instead
by $\kappa_{\bigoplus}$ and is given by
\begin{eqnarray}
\kappa_{\bigoplus} &=&
\frac{\kappa_{\ast}}{1+f}\frac{2+2f+4\left(\frac{df}{d\sqrt{\kappa_{\ast}}\varphi}\right)^2}
{2+2f+3\left(\frac{df}{d\sqrt{\kappa_{\ast}}\varphi}\right)^2}
\end{eqnarray}
where $\sqrt{\kappa_{\ast}}$ is added to make
$\sqrt{\kappa_{\ast}}\varphi$ dimensionless, which is the
convention we shall always follow below. $\kappa_{\bigoplus}$
itself is obviously not a constant and we measure only it present-day value, $\kappa _{\bigoplus 0}$.

Varying the action with respect to the scalar field, $\varphi$,
gives the scalar field equation of motion
\begin{eqnarray}\label{eq:sfeom}
\nabla^{a}\nabla_{a}\varphi + \frac{\partial
V(\varphi)}{\partial\varphi} +
\frac{R}{2\kappa_{\ast}}\frac{\partial
f(\varphi)}{\partial\varphi} &=& 0.
\end{eqnarray}

Since we will follow the motions of dark matter particles in the
$N$-body simulations, so we also need their geodesic
equation. The dark-matter Lagrangian for a point particle with
mass $m_{0}$ is
\begin{equation}\label{eq:DMLagrangian}
\mathcal{L}_{\mathrm{CDM}}(\mathbf{y})=-\frac{m_{0}}{\sqrt{-g}}\delta (%
\mathbf{y}-\mathbf{x}_{0})\sqrt{g_{ab}\dot{x}_{0}^{a}\dot{x}_{0}^{b}},
\end{equation}%
where $\mathbf{y}$ is the general coordinate and $\mathbf{x}_{0}$
is the coordinate of the centre of the particle. From this
equation we derive the corresponding energy-momentum tensor:
\begin{equation}\label{eq:DMEMT_particle}
T_{\mathrm{CDM}}^{ab}=\frac{m_{0}}{\sqrt{-g}}\delta (\mathbf{y}-\mathbf{x}%
_{0})\dot{x}_{0}^{a}\dot{x}_{0}^{b}.
\end{equation}
Taking the conservation equation for dark matter particles (which,
unlike in \citep{Li:2009sy, Li:2010}, does not couple to any other
matter species, including the scalar field $\varphi$), the
geodesic equation follows as usual:
\begin{eqnarray}\label{eq:geodesic}
\ddot{x}^{a}_{0} + \Gamma^{a}_{bc}\dot{x}^{b}_{0}\dot{x}^{c}_{0}
&=& 0,
\end{eqnarray}
where the second term on the left-hand side accounts for gravity.

Eqs.~(\ref{eq:Einstein}, \ref{eq:sfeom} , \ref{eq:geodesic})
contain all the physics needed for the following
analysis, though certain approximations and simplifications might
have to be made in due course to make direct connection to
$N$-body simulations.

We will consider an inverse power-law potential for the scalar
field,
\begin{eqnarray}\label{eq:potential}
V(\varphi) &=&
\frac{\Lambda^{4}}{\left(\sqrt{\kappa_{\ast}}\varphi
\right)^{\alpha}},
\end{eqnarray}
where $\alpha$ is a dimensionless constant and $\Lambda $ is a
constant with dimensions of mass. This potential has also been
adopted in various background or linear perturbation studies of
scalar fields (either minimally or non-minimally coupled); the
tracking behaviour its produces makes it a good dark energy
candidate and for that purpose we shall choose $\alpha \sim
\mathcal{O}(0.1-1)$. Meanwhile, the coupling between the scalar
field and the curvature tensor is chosen to be a non-minimal one:
\begin{eqnarray}\label{eq:coupling_function}
f(\varphi) &=& \gamma\kappa_{\ast}\varphi^2,
\end{eqnarray}
where $\gamma$ is another dimensionless constant characterising
the strength of the coupling. Note that here again $\kappa_\ast$
is added into $f(\varphi)$ and $V(\varphi)$ to make a dimensionless
quantity $\sqrt{\kappa_{\ast}}\varphi$. Although the exact value
of $\kappa_{\ast}$ is unknown, so is $\varphi$ and we can
solve for $\sqrt{\kappa_{\ast}}\varphi $ instead of $\varphi $, not caring
about the exact individual values of $\sqrt{\kappa_{\ast }}$ and $\varphi $.

\subsection{The Non-Relativistic Limits}

\label{subsect:nonrel}

The $N$-body simulation only probes the motion of particles at
late times, and we are not interested in extreme conditions such
as black hole formation and evolution, so we can take the
non-relativistic limit of the above equations as a good
approximation.

The existence of the scalar field and its coupling to the
curvature leads to several possible changes with respect to the
$\Lambda$CDM paradigm:

\begin{enumerate}
\item The scalar field has its own energy-momentum tensor, which
could change the source term of the Poisson equation because the
scalar field, unlike the cosmological constant, can cluster
(though the clustering is often quite weak in scalar field
models). Also, unlike in
coupled scalar field models, here the $\vec{\nabla}^2\varphi$ term
will appear in the Poisson equation.

\item The background cosmic expansion rate is in general modified,
and can either slow down or speed up the rate of structure
formation.

\item The two gravitational potentials in the conformal Newtonian
gauge metric
$ds^2=a^2(1+2\phi)d\tau^2+a^2(1-2\psi)\delta_{ij}dx^{i}dx^j$, in
which $\tau$ and $x^{i}$ are respectively the conformal time and
comoving coordinate, are no longer equal to each other (as in
general relativity), but are instead related by
$\vec{\nabla}^{2}\varphi$ (see below).
\end{enumerate}

It therefore becomes clear that the following two equations, in
their non-relativistic forms, need to be solved in order to obtain
the gravitational force on particles:

\begin{enumerate}
\item The scalar field equation of motion, which is used to
compute \emph{explicitly} the value of the scalar field $\varphi$
at any given time and position;

\item The Poisson equation, which is used to determine the
gravitational potential and force at any given time
and position from the local energy density and pressure, which
includes the contribution from the scalar field (obtained from the
$\varphi$ equation of motion).
\end{enumerate}
Note that unlike in the coupled scalar field models, there is no
fifth force because there is no direct coupling to the particles.
The scalar coupling to the curvature, however, does modify the
gravitational potential so that gravity no longer follows
Einstein's prescription and so this is a
modified gravity theory.

We now describe these two equations in turn. For the scalar field
equation of motion, we denote by $\bar{\varphi}$ the background
value of $\varphi $ and write $\delta \varphi \equiv \varphi
-\bar{\varphi}$. Then using the expressions given in Appendix
\ref{appen:expression} we write
\begin{eqnarray}
a^{2}\nabla^{a}\nabla_{a}\varphi &=& \varphi'' +
2\mathcal{H}\varphi' + \vec{\nabla}_{\mathbf{x}}^{2}\varphi -
2\phi\varphi''\nonumber\\
&& - \left(\phi'+3\psi'+4\mathcal{H}\phi\right)\varphi'
\end{eqnarray}
in which $'=d/d\tau$ with $\tau$ the conformal time,
$\vec{\nabla}_{\mathbf{x}}$ is the derivative with respect to the
comoving coordinate $\mathbf{x}$, and $\mathcal{H}=a'/a$. Then,
with the background part subtracted, Eq.~(\ref{eq:sfeom}) can be
rewritten as
\begin{eqnarray}
\delta\varphi'' + 2\mathcal{H}\delta\varphi' +
\vec{\nabla}_{\mathbf{x}}^{2}\delta\varphi +
\left[V_{,\varphi}(\varphi)-V_{,\varphi}(\bar{\varphi})\right]a^2\nonumber\\
-2\phi\bar{\varphi}''-\left(\phi'+3\psi'+4\mathcal{H}\phi\right)\bar{\varphi}'
\nonumber\\ +
\frac{1}{2\kappa_{\ast}}\left[Rf_{\varphi}(\varphi)-\bar{R}f_{\varphi}(\bar{\varphi})\right]a^2
&=& 0,\nonumber
\end{eqnarray}
in which a bar denotes the background value, and the subscript
$_\varphi$ denotes derivatives with respect to $\varphi$. Note that
$\vec{\nabla}^{2}_{\mathbf{x}}$ has the same sign as
$\vec{\nabla}^{2}_{\mathbf{r}}$.

In our $N$-body simulations we shall work in the quasi-static
limit, \emph{i.e.}, we assume that the spatial gradients are much
greater than the time derivatives,
$|\vec{\nabla}_{\mathbf{x}}\varphi |\gg |\frac{\partial \delta
\varphi }{\partial \tau}|$. Therefore, the time derivatives in the
above equation are dropped and we obtain the simplified version
\begin{eqnarray}\label{eq:WFphiEOM0}
&&c^{2}\vec{\partial}_{\mathbf{x}}^{2}(a\delta\varphi)\\ &=&
a^{3}\left[V_{\varphi}(\varphi)-V_{\varphi}(\bar{\varphi})\right]
+ \frac{1}{2\kappa_\ast}\left[Rf_{\varphi}(\varphi)
-\bar{R}f_\varphi(\bar{\varphi})\right]a^3,\nonumber
\end{eqnarray}
in which $\vec{\partial}
_{\mathbf{x}}^{2}=-\vec{\nabla}_{\mathbf{x}}^{2}=+\left(\partial
_{x}^{2}+\partial _{y}^{2}+\partial _{z}^{2}\right) $ due to our
sign convention $(+,-,-,-)$, and we have restored the factor
$c^{2}$ in front of $\vec{\nabla}_{\mathbf{x}}^{2}$ (the $\varphi
$ here and in the remaining of this paper is $c^{-2}$ times the
$\varphi$ in the original Lagrangian unless otherwise stated).
Note that here $V$ has the dimension of \emph{mass} density rather
than \emph{energy} density.

To complete Eq.~(\ref{eq:WFphiEOM0}), we still need expressions
for $R$ and $\bar{R}$, which are again obtained using the
quantities in Appendix \ref{appen:expression}:
\begin{eqnarray}
R &=& -\frac{6}{a^2}\frac{a''}{a}(1-2\phi) \nonumber\\ && +
\frac{1}{a^2}\left[6\psi''+6\mathcal{H}\left(\phi'+3\psi'\right)
-4\vec{\partial}_{\mathbf{x}}^2\psi+2\vec{\partial}_{\mathbf{x}}^2\phi\right],\
\ \ \\
\bar{R} &=& -\frac{6}{a^2}\frac{a''}{a}
\end{eqnarray}
and so
\begin{eqnarray}\label{eq:aid1}
Rf_{\varphi}-\bar{R}\bar{f}_{\varphi} &\doteq&
\bar{f}_{\varphi}\delta R + \bar{R}\delta f_{\varphi}\nonumber\\
&\doteq&
-\frac{1}{a^2}\bar{f}_{\varphi}\left(4\vec{\partial}_{\mathbf{x}}^2\psi
-2\vec{\partial}_{\mathbf{x}}^2\phi\right) -
\frac{6}{a^2}\frac{a''}{a}\delta f_{\varphi}\ \ \
\end{eqnarray}
where we have again dropped time derivatives of $\phi$ and $\psi$
since they are small compared with the corresponding spatial
gradients, and $\delta R\equiv R-\bar{R}$, $\delta f_\varphi
\equiv f_\varphi-\bar{f}_\varphi$.

Since only $\phi$ but not $\psi$ appears in the
Poisson equation (shown below) , we also want to eliminate the $\psi$
in the scalar field equation of motion. This is easy in general relativity,
because there we have the simple relation $\phi =\psi$, which
unfortunately no longer holds in scalar-tensor theories. However, we
could use the $i-j$ components of the Einstein equation $G_{\ j}^{i}=\kappa_{\ast }T_{\ j}^{i}$ ($i\neq j$) to get a new relation between $\phi $ and 
$\psi$. Noting that our $N$-body simulations probe the very late time
evolution (when radiation is negligible) when the only significant
source for $T_{\ j}^{i}$ ($i\neq j$) is the scalar field, and 
\begin{eqnarray}
\nabla^{i}\nabla_{j}f &=& -\frac{1}{a^2}\partial_i\partial_jf\ \
(i\neq j)
\end{eqnarray}
where $\partial_i\equiv\partial/\partial x^i$, we could write the
$i-j$ component of Einstein equation as
\begin{eqnarray}
\partial_i\partial_j\left(\phi-\psi\right) &=&
-\frac{c^2}{1+f}\partial_i\partial_jf\nonumber
\end{eqnarray}
which gives approximately
\begin{eqnarray}
\partial_i\left(\phi-\psi\right) &=&
-\frac{c^2}{1+f}\partial_if\nonumber
\end{eqnarray}
and so
\begin{eqnarray}\label{eq:aid2}
4\vec{\partial}_{\mathbf{x}}^2\psi -
2\vec{\partial}_{\mathbf{x}}^2\phi &\doteq&
2\vec{\partial}_{\mathbf{x}}^2\phi +
\frac{4}{1+f}\vec{\partial}_{\mathbf{x}}^2f\nonumber\\
&\doteq& 2\vec{\partial}_{\mathbf{x}}^2\phi +
\frac{4\bar{f}_{\varphi}}{1+\bar{f}}c^2\vec{\partial}_{\mathbf{x}}^2\delta\varphi.
\end{eqnarray}
It is important to note that in the second line of Eq.~(\ref{eq:aid2}) we
have implicitly linearised the equation; this is valid only if 
$f(\varphi)$ is not strongly nonlinear and $|\delta \varphi /\varphi |\ll 1$. 
It turns out that the model considered in this work satisfies these
criteria ($f(\varphi)\propto\varphi^{2}$). If either $V(\varphi)$ or $f(\varphi )$ 
is highly nonlinear, then we might have $|\delta \varphi |\sim
\varphi$; in that case we should not approximate $f$ to $\bar{f}$ even in the coefficients of the perturbation variables such as 
$\vec{\partial}_{\mathbf{x}}^{2}\delta \varphi $ here, or write $\vec{\partial}_{\mathbf{x}}^{2}f=\bar{f}_{\varphi }\vec{\partial}_{\mathbf{x}}^{2}\delta \varphi$. 
The reason for the latter stricture is as
follows: if $f(\varphi)$ is highly nonlinear, then $f$ might change a lot
even if $\varphi$ fluctuates a little, implying that for the linearisation
to apply on our spatial grid we need very small grid sizes which are
impossible; moreover, it becomes complicated to decide which
solution we should linearise around, as the values of $f$ in that
area which we look at might be very different from the background value $\bar{f}$. 
The strategy for this situation is simple: instead of
writing $\vec{\partial}_{\mathbf{x}}^{2}f=\bar{f}_{\varphi }\vec{\partial}_{\mathbf{x}}^{2}\delta \varphi$, we difference $f(\varphi )$ directly, 
because we know the value of $f(\varphi)$ in every grid cell. This
will ensure no linearisation error. In what follows, however, we shall use
Eq.~(\ref{eq:aid2}), which causes negligible linearisation error but
simplifies the equations a lot. We shall also write $f\doteq \bar{f}$ in the
coefficients of perturbation quantities such as $\vec{\partial}_{\mathbf{x}}^{2}\delta \varphi $ and $\vec{\partial}_{\mathbf{x}}^{2}\Phi$.

Substituting Eqs.~(\ref{eq:aid1}, \ref{eq:aid2}) into
Eq.~(\ref{eq:WFphiEOM0}) and rearranging, we complete the
derivation of the scalar field equation of motion in the weak
field limit, ending up with
\begin{eqnarray}
&&\left[1+\frac{2\bar{f}^{2}_{\varphi}}{\kappa_\ast(1+\bar{f})}\right]
c^{2}\vec{\partial}_{\mathbf{x}}^{2}\left(a\delta\varphi\right)\nonumber\\
&=& a^{3}\left[V_\varphi(\varphi)-V_\varphi(\bar{\varphi})\right]
-
\frac{\bar{f}_{\varphi}}{2\kappa_\ast}\vec{\partial}_{\mathbf{x}}^{2}\Phi
-\frac{3}{\kappa_{\ast}}\frac{a''}{a}a\delta f_{\varphi}
\end{eqnarray}
for our general Lagrangian Eq.~(\ref{eq:Lagrangian}) and
\begin{eqnarray}\label{eq:WFphiEOM}
&&\left[1+\frac{8\gamma^2\kappa_{\ast}\bar{\varphi}^{2}}{1+\gamma\kappa_{\ast}\bar{\varphi}^2}\right]
c^{2}\vec{\partial}_{\mathbf{x}}^{2}\left(a\sqrt{\kappa_{\ast}}\delta\varphi\right)\nonumber\\
&=&
-\gamma\sqrt{\kappa_{\ast}}\bar{\varphi}\vec{\partial}_{\mathbf{x}}^{2}\Phi
-
6\gamma\left(\mathcal{H}'+\mathcal{H}^2\right)(a\sqrt{\kappa_{\ast}}\delta\varphi)\nonumber\\
&& -
\alpha\kappa_{\ast}\Lambda^4a^{3}\left[\frac{1}{\left(\sqrt{\kappa_{\ast}}\varphi\right)^{1+\alpha}}
-
\frac{1}{\left(\sqrt{\kappa_{\ast}}\bar{\varphi}\right)^{1+\alpha}}\right]
\end{eqnarray}
for the model specified by Eqs.~(\ref{eq:potential},
\ref{eq:coupling_function}), where $\Phi\equiv a\phi$.

Next consider the Poisson equation, which is obtained from the
Einstein equation in the weak-field and slow-motion limits. Here
we use the $0-0$ component of the Ricci curvature tensor, which is
given as
\begin{eqnarray}
R^{0}_{\ 0} &=&
-3\left(\frac{a''}{a}-\mathcal{H}^2\right)(1-2\phi)\nonumber\\ &&
+ 3\psi'' + 3\mathcal{H}\left(\psi'+\phi'\right) +
\vec{\partial}_{\mathbf{x}}^{2}\phi
\end{eqnarray}
using the expressions in Appendix~\ref{appen:expression}.
According to the Einstein equations,
\begin{eqnarray}\label{eq:EinsteinEqn}
R^{0}_{\ 0} &=&
\frac{\kappa_{\ast}}{2}(\rho_{\mathrm{TOT}}+3p_{\mathrm{TOT}})a^2
\end{eqnarray}
where $\rho _{\mathrm{TOT}}$ and $p_{\mathrm{TOT}}$ are the total
energy density and pressure, respectively. Using these two
equations and subtracting the background part (which is just the Raychaudhuri equation), it is straightforward to find that
\begin{eqnarray}
\vec{\partial}_{\mathbf{x}}^{2}\Phi &=&
\frac{\kappa_{\ast}a^{3}}{2}\left[(\rho_{\mathrm{TOT}}+3p_{\mathrm{TOT}})
- (\bar{\rho}_{\mathrm{TOT}}+3\bar{p}_{\mathrm{TOT}})\right].\ \ \
\end{eqnarray}
in which we have dropped terms involving time derivatives of
$\psi, \phi$ and $\mathcal{H}^2\phi$, because they are
much smaller than $\vec{\partial}^{2}_{\mathbf{x}}\phi$ in the
quasi-static limit. Using the energy-momentum tensor expressed in
Eq.~(\ref{eq:Einstein}), the above equation can be rewritten as
\begin{eqnarray}
&&\vec{\partial}_{\mathbf{x}}^{2}\Phi\\ &\doteq&
\frac{\kappa_{\ast}a^3}{2}\bar{\rho}_{m}\left(\frac{\delta}{1+f}-\frac{1}{1+\bar{f}}\right)
-
\frac{\bar{f}_{\varphi}}{2(1+\bar{f})}c^{2}\vec{\partial}^{2}_{\mathbf{x}}\left(a\delta\varphi\right)\nonumber\\
&&
-\kappa_{\ast}a^{3}\left[\frac{V(\varphi)}{1+f}-\frac{V(\bar{\varphi})}{1+\bar{f}}\right]\nonumber\\
&&+a\left(\frac{1}{1+f}-\frac{1}{1+\bar{f}}\right)\left(\kappa_{\ast}\bar{\varphi}'^{2}+\frac{3}{2}f''\right)\nonumber
\end{eqnarray}
for the general Lagrangian Eq.~(\ref{eq:Lagrangian}) and
\begin{eqnarray}\label{eq:WFPoisson}
&&\vec{\partial}_{\mathbf{x}}^{2}\Phi\\ &=&
\frac{3}{2}\left(1+\gamma\kappa_{\ast}\bar{\varphi}_0^2\right)\Omega_mH_0^2
\left[\frac{\delta}{1+\gamma\kappa_{\ast}\varphi^2}-\frac{1}{1+\gamma\kappa_\ast\bar{\varphi}^2}\right]\nonumber\\
&&-\frac{\gamma\sqrt{\kappa_{\ast}}\bar{\varphi}}{1+\kappa_{\ast}\bar{\varphi}^2}c^{2}
\vec{\partial}_{\mathbf{x}}^{2}\left(a\sqrt{\kappa_{\ast}}\delta\varphi\right)\nonumber\\
&&-\left[\frac{\kappa_{\ast}\Lambda^4a^3}{\left(1+\gamma\kappa_{\ast}\varphi^2\right)
\left(\sqrt{\kappa_{\ast}}\varphi\right)^\alpha}-
\frac{\kappa_{\ast}\Lambda^4a^3}{\left(1+\gamma\kappa_{\ast}\bar{\varphi}^2\right)
\left(\sqrt{\kappa_{\ast}}\bar{\varphi}\right)^\alpha}\right]\nonumber\\
&&+\left[(1+3\gamma)\kappa_{\ast}\bar{\varphi}'^2+3\gamma\kappa_{\ast}\bar{\varphi}\bar{\varphi}''\right]a\nonumber\\
&&\ \ \ \ \times\left[\frac{1}{1+\gamma\kappa_{\ast}\varphi^2}
-\frac{1}{1+\gamma\kappa_{\ast}}\bar{\varphi}^2\right]\nonumber
\end{eqnarray}
for the model specified by Eqs.~(\ref{eq:potential},
\ref{eq:coupling_function}). In these equations $\bar{\rho}_{m}$ is
the background density for matter,
$\delta\equiv\rho_m/\bar{\rho}_m$, and we have used the definition
of $\Omega_m$ given in Appendix~\ref{appen:bkgd}. We have also neglected the contribution from
$\dot{\delta\varphi}$ to the total density and pressure, because
in the quasi-static limit we have
$|\delta\varphi''|\ll|\vec{\partial}^{2}_{\mathbf{x}}\delta\varphi|$
and
$\delta\varphi'^{2}\lesssim|\bar{\varphi}'\delta\varphi'|\lesssim|\mathcal{H}\delta\varphi'|\ll\vec{\partial}^{2}_{\mathbf{x}}\delta\varphi|$
(which is confirmed by the $N$-body simulation
results\footnotemark[1]).

\footnotetext[1]{According to Eq.~(\ref{eq:WFphiEOM}) we have
$\vec{\partial}_{\mathbf{x}}^{2}\left(a\sqrt{\kappa_{\ast}}\delta\varphi\right)
\sim\mathcal{O}\left(\vec{\partial}^{2}_{\mathbf{x}}\Phi\right)$,
implying that
$a\sqrt{\kappa_{\ast}}\delta\varphi\sim\mathcal{O}(\Phi)$, so
neglecting time derivatives of $\delta\varphi$ is just like
dropping time derivatives of $\psi$ and $\phi$, which we have
already done to obtain the modified Poisson equation.}

Finally, the equation of motion of the dark matter particles is the same as in general relativity
\begin{eqnarray}
\ddot{\mathbf{x}} + 2\frac{\dot{a}}{a}\dot{\mathbf{x}} &=&
-\frac{1}{a^{3}}\vec{\nabla}_{\mathbf{x}}\Phi
\end{eqnarray}
in which $\Phi$ is determined by the modified Poisson equation
Eq.~(\ref{eq:WFPoisson}). The canonical momentum conjugate to
$\mathbf{x}$ is $\mathbf{p}=a^{2}\dot{\mathbf{x}}$ so from the
equation above we have
\begin{eqnarray}\label{eq:WFdxdtcomov}
\frac{d\mathbf{x}}{dt} &=& \frac{\mathbf{p}}{a^{2}},\\
\label{eq:WFdpdtcomov} \frac{d\mathbf{p}}{dt} &=&
-\frac{1}{a}\vec{\nabla}_{\mathbf{x}}\Phi.
\end{eqnarray}

Eqs.~(\ref{eq:WFphiEOM}, \ref{eq:WFPoisson}, \ref{eq:WFdxdtcomov},
\ref{eq:WFdpdtcomov}) will be used in the code to evaluate the
forces on the dark-matter particles and evolve their positions and
momenta in time. But before applying them to the code we still
need to switch to code units (see Sect.~\ref{subsect:codeunit}),
further simplify them and create the discrete version (see
Appendix~\ref{appen:discret}).

\subsection{Code Units}

\label{subsect:codeunit}

In our numerical simulation we use a modified version of
\texttt{MLAPM} (\citep{Knebe:2001}), and we will have to change or
add our Eqs.~(\ref{eq:WFphiEOM}, \ref{eq:WFPoisson},
\ref{eq:WFdxdtcomov}, \ref{eq:WFdpdtcomov}) to it. The first
step is to convert the quantities to the code units of
\texttt{MLAPM}. Here, we briefly summarise the main results.

The (modified) \texttt{MLAPM} code uses the following internal
units (where a subscript $_{c}$ stands for "code"):
\begin{eqnarray}
\mathbf{x}_{c} &=& \mathbf{x}/B,\nonumber\\
\mathbf{p}_{c} &=& \mathbf{p}/(H_{0}B)\nonumber\\
t_{c} &=& tH_{0}\nonumber\\
\Phi_{c} &=& \Phi/(H_{0}B)^{2}\nonumber\\
\rho_{c} &=& \rho/\bar{\rho},\nonumber\\
u &=& ac^{2}\sqrt{\kappa}\delta\varphi/\left(H_0B\right)^2,
\end{eqnarray}
where $B$ denotes the comoving size of the simulation box, $H_{0}$
is the present Hubble constant, and $\rho$ is the matter density.
In the last line the quantity $u$ is the scalar field
\emph{perturbation} $\delta\varphi$ expressed in terms of code
units and is new to the \texttt{MLAPM} code.

In terms of $u$, as well as the (dimensionless) background value
of the scalar field, $\sqrt{\kappa }\bar{\varphi}$, some relevant
quantities are expressed in full as
\begin{eqnarray}
V(\varphi) &=&
\frac{\Lambda^{4}}{\left(\sqrt{\kappa}\bar{\varphi}+\frac{B^2H^2_0}{ac^{2}}u\right)^{\alpha}},\nonumber\\
f(\varphi) &=&
1+\gamma\left(\sqrt{\kappa}\bar{\varphi}+\frac{B^2H^2_0}{ac^{2}}u\right)^2,\nonumber\\
V_{\varphi}(\varphi) &=&
-\alpha\frac{\sqrt{\kappa}\Lambda^{4}}{\left(\sqrt{\kappa}\bar{\varphi}+\frac{B^2H^2_0}{ac^{2}}u\right)^{1+\alpha}},\nonumber\\
f_{\varphi}(\varphi) &=&
2\gamma\sqrt{\kappa}\left(\sqrt{\kappa}\bar{\varphi}+\frac{B^2H^2_0}{ac^{2}}u\right),
\end{eqnarray}
and the background counterparts of these quantities can be
obtained simply by setting $u=0$ (recall that $u$ represents the
perturbed part of the scalar field) in the above equations.

We also define
\begin{eqnarray}\label{eq:lambda}
\lambda &\equiv& \frac{\kappa\Lambda^4}{3H_{0}^{2}},
\end{eqnarray}
which will be used frequently below.

Making discrete versions of the above equations for $N$-body
simulations is then straightforward, and we refer the interested
readers to Appendix~\ref{appen:discret} to the whole treatment,
with which we can now proceed to do $N$-body simulations.

\begin{figure*}[tbp]
\centering \includegraphics[scale=1.05] {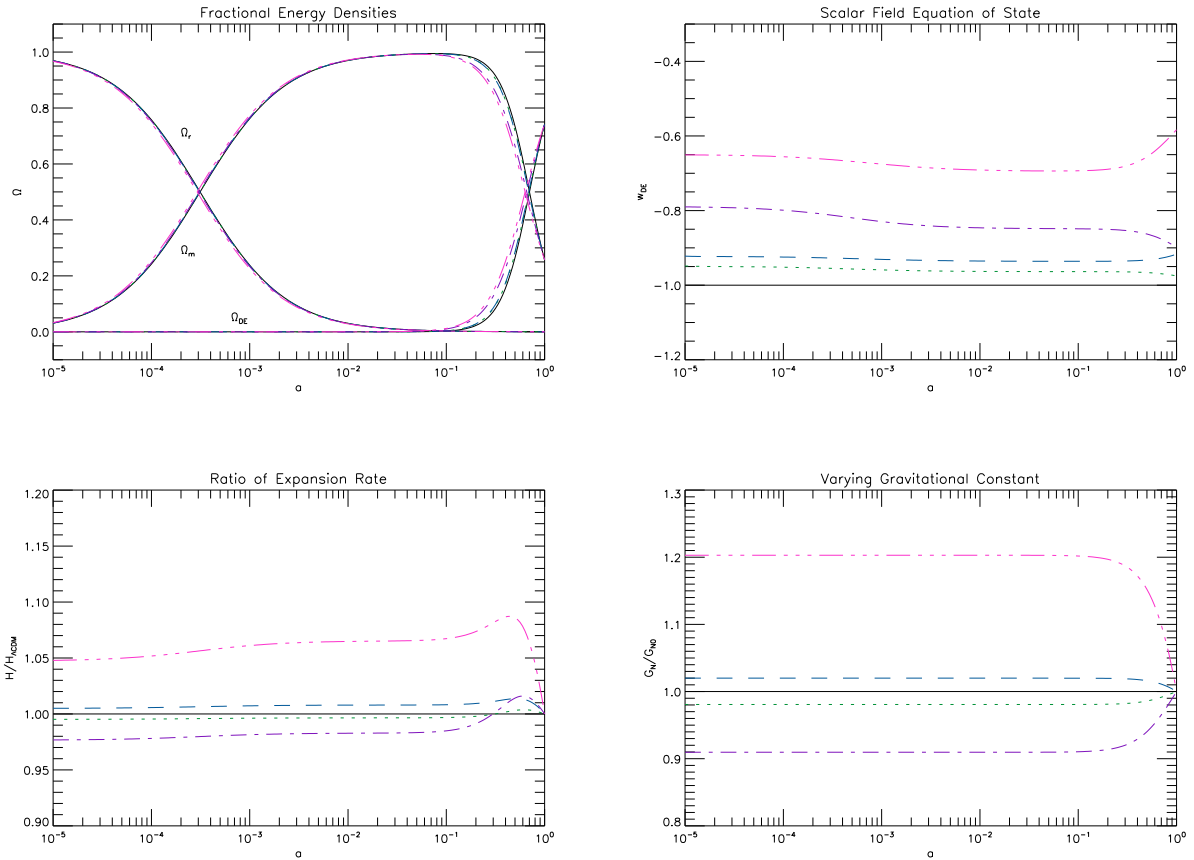}
\caption{(Color Online) The background evolution in the extended
quintessence models. \emph{Upper-left panel}: the fractional
energy densities for matter ($\Omega_m$), radiation ($\Omega_r$)
and the scalar field dark energy ($\Omega_{\mathrm{DE}}$), as
indicated besides the curves, as functions of the scale factor $a$
($a_0=1$ today). \emph{Upper-right panel}: the scalar field
equation of state $w=p_{\mathrm{DE}}/\rho_{\mathrm{DE}}$ as a
function of $a$. \emph{Lower-left panel}: the ratio between the
Hubble expansion rates of the extended quintessence model and
$\Lambda$CDM as a function of $a$. \emph{Lower-right panel}: the
$a$-evolution of the effective gravitational constant that governs
the growth of matter density perturbations ($G_{\mathrm{N}0}$ is
its value today). In all panels the black solid, green dot, blue
dashed, purple dot-dashed, pink dot-dot-dot-dashed curves
represent respectively the results for $\Lambda$CDM and extended
quintessence models with $(\alpha, \gamma)=(0.1,-0.2)$,
$(0.1,0.2)$, $(0.5,-0.2)$, $(0.5,0.2)$.} \label{fig:background}
\end{figure*}

\begin{figure*}[tbp]
\centering \includegraphics[scale=1.85] {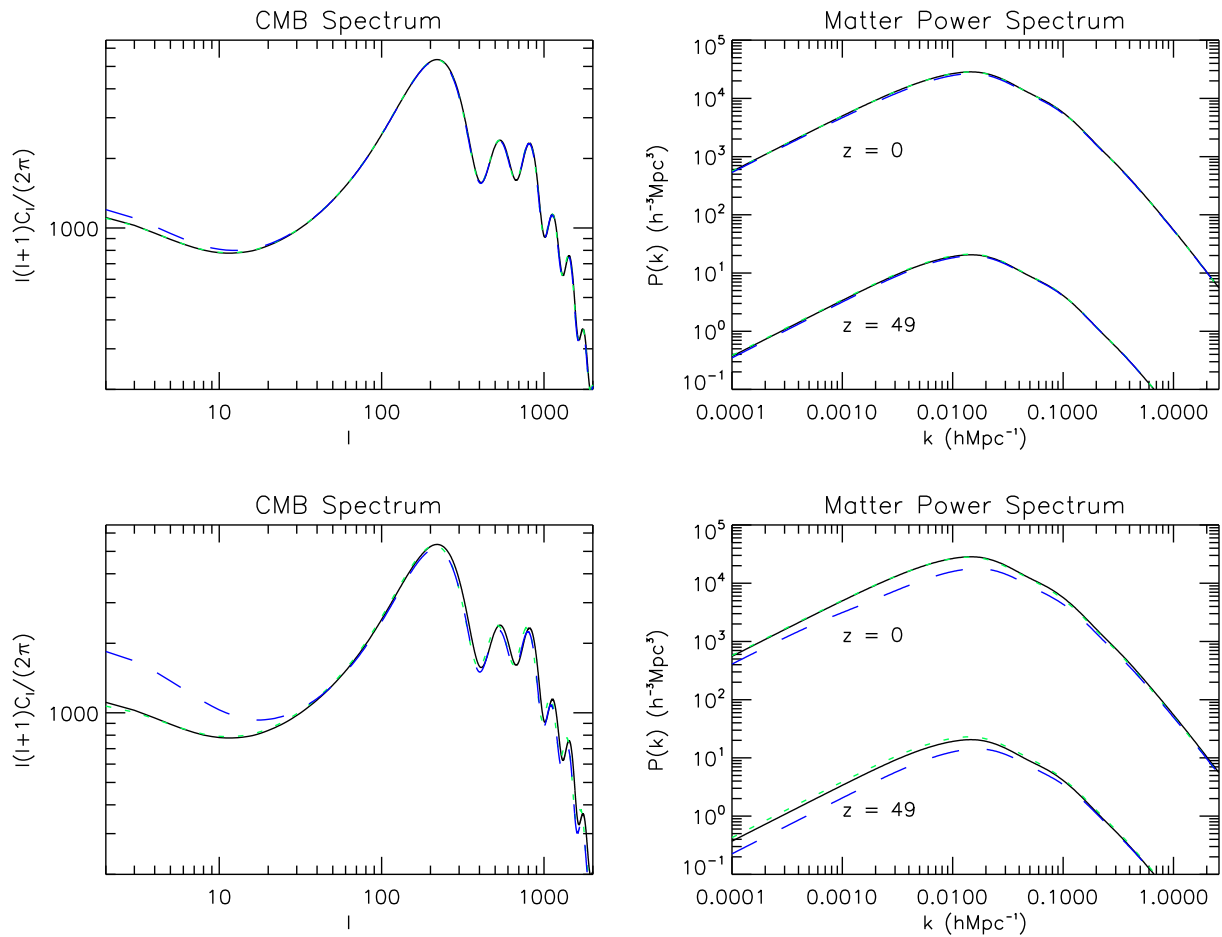}
\caption{(Color Online) The CMB (\emph{left panels}) and matter
power spectra (\emph{right panels}) for the extended quintessence
models compared with those of the $\Lambda$CDM. The upper panels
are for the models with $\alpha=0.1$ while the lower panels are
for those with $\alpha=0.5$. The black solid, green dotted and
blue dashed curves represent respectively the curves for
$\Lambda$CDM and extended quintessence with $\gamma=-0.2$ and
$\gamma=0.2$. For the matter power spectra, we plot the results
for two different output redshifts, $z=0$ and $49$, as indicated
below the curves.} \label{fig:linpert}
\end{figure*}

\begin{figure*}[tbp]
\centering \includegraphics[scale=0.85] {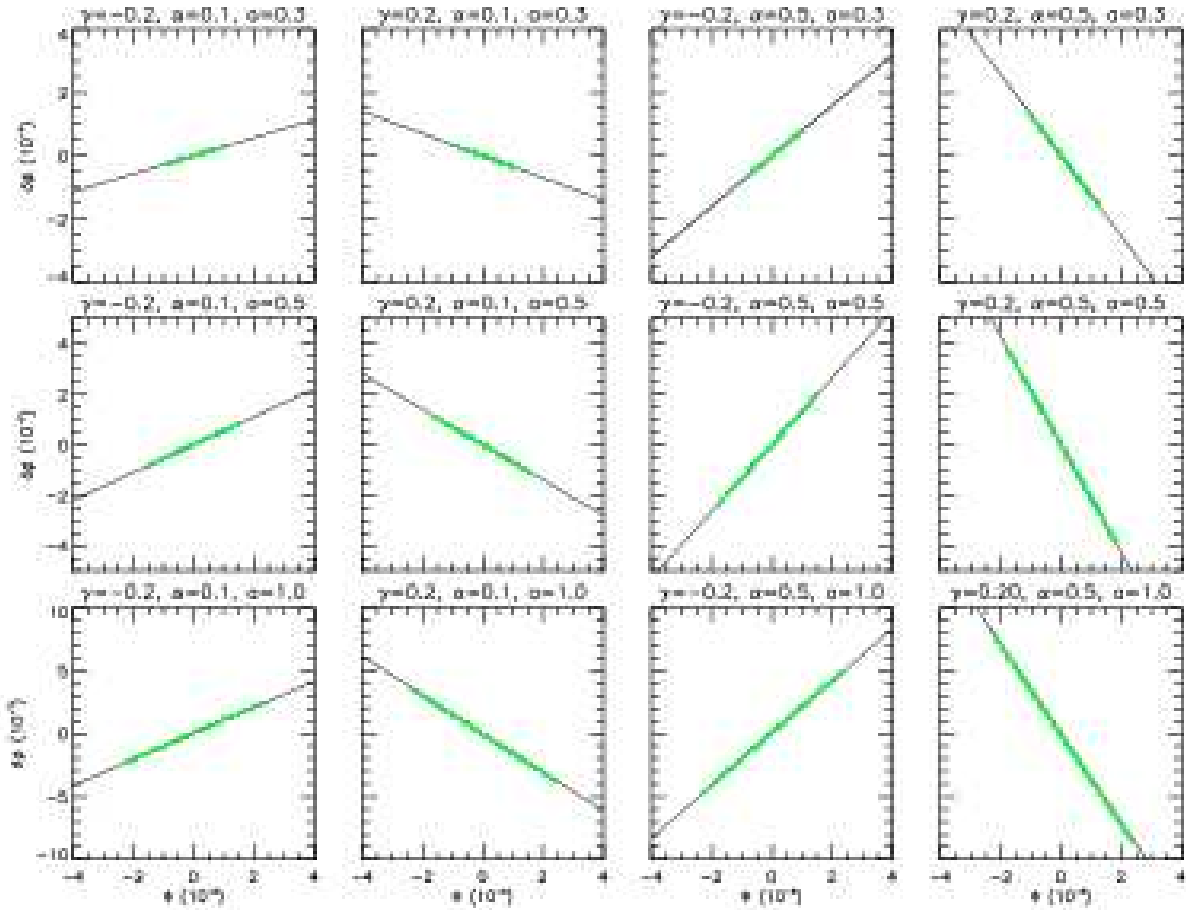}
\caption{(Color Online) The relation between the magnitudes of the
scalar field perturbation $a\sqrt{\kappa_{\ast}\delta\varphi}$ (in
unit of $10^{-7}$) and gravitational potential $\Phi$ (in unit of
$10^{-6}$) for the four extended quintessence models (the four
columns) at three different output times $a=0.3, 0.5, 1.0$ (the
three rows) as indicated above the frames. The black solid line in
each panel represents the analytical approximation
Eq.~(\ref{eq:aid3}) (see text) and the $\sim10,000$ green dots the
results from a thin slice of our simulation boxes.}
\label{fig:scal_pot}
\end{figure*}

\begin{figure*}[tbp]
\centering \includegraphics[scale=1.6] {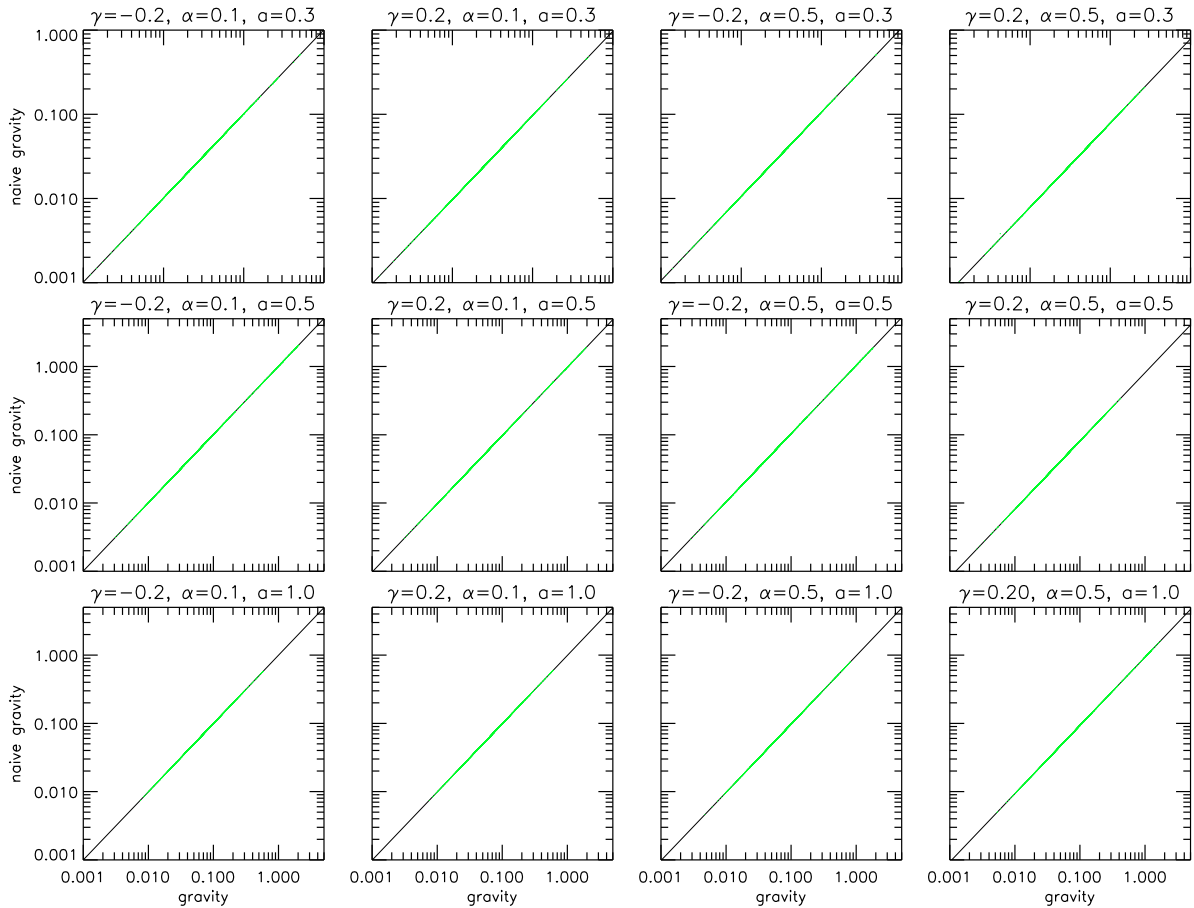} \caption{(Color
Online) The relation between the magnitudes of the na\"{\i}ve
gravity (see text) and full gravity for the four extended
quintessence models (the four columns) at three different output
times $a=0.3, 0.5, 1.0$ (the three rows) as indicated above the
frames. The black solid line in each panel represents the
analytical approximation Eq.~(\ref{eq:G_N}) (see text) and the
$\sim10,000$ green dots the results from the simulations.}
\label{fig:force}
\end{figure*}

\begin{figure*}[tbp]
\centering \includegraphics[scale=1.8] {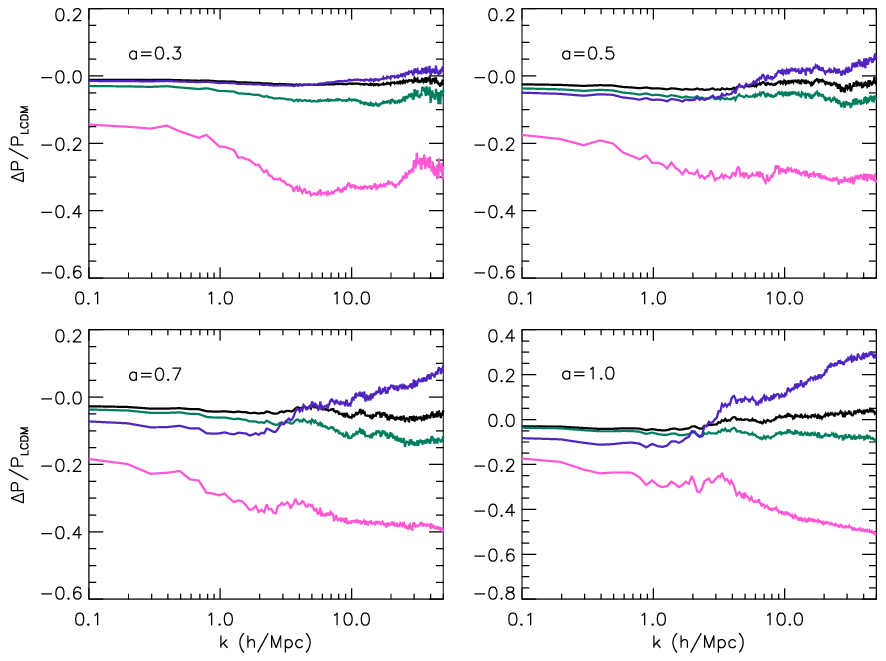} \caption{(Color
Online) The fractional difference between the nonlinear power
spectra for the extended quintessential and $\Lambda$CDM models.
The results for the four models of $(\alpha,\gamma)=(0.1,-0.2)$,
$(0.1,0.2)$, $(0.5,-0.2)$ and $(0.5,0.2)$ are respectively
represented by the black, green, purple and pink curves. The four
panels are for four output times $a=0.3$, $0.5$, $0.7$ and $1.0$
as indicated inside the corresponding frames.} \label{fig:power}
\end{figure*}

\section{Simulation Details}

\label{sect:simu}

\subsection{The $N$-Body Code}

\label{subsect:code}

Some of our main modifications to the \texttt{MLAPM} code for the
coupled scalar field model are:

\begin{enumerate}
\item We have added a solver for the scalar field, based on
Eq.~(\ref{eq:u_phi_EOM}). It uses a nonlinear Gauss-Seidel scheme
for the relaxation iteration and the same criterion for
convergence as the default Poisson solver. But it adopts a V-cycle
instead of the self-adaptive scheme in arranging the Gauss-Seidel
iterations.

\item The value of $u$ solved in this way is then used to calculate
the total matter density, which completes the calculation of the
source term for the Poisson equation. The latter is then solved
using a fast Fourier transform on the domain grids and self-adaptive
Gauss-Seidel iteration on refinements.

\item The gravitational potential $\Phi$ obtained in this way is
then used to compute the force, which is used to displace and kick
the particles.
\end{enumerate}

There are a lot of additions and modifications to ensure smooth
interface and the newly added data structures. For the output, as
there are multilevel grids all of which host particles, the
composite grid is inhomogeneous and so we choose to output the
positions and momenta of the particles, plus the gravity and
values of $\Phi$ and $u$ at the positions of these particles. We
also output the potential and scalar field values on the $128^{3}$
domain grid.

\subsection{Physical and Simulation Parameters}

\label{subsect:param}

The physical parameters we use in the simulations are as follows:
the present-day dark-energy fractional energy density
$\Omega_{\mathrm{DE}}=0.743$ and $\Omega _{m}=\Omega
_{\mathrm{CDM}}+\Omega _{\mathrm{B}}=0.257$,
$H_{0}=71.9$~$km/s/Mpc$, $n_{s}=0.963$, $\sigma_{8}=0.769$. Our
simulation box has a size of $64h^{-1}$~Mpc, where
$h=H_{0}/(100~\mathrm{km/s/Mpc})$. We simulate four models, with
parameters $(\alpha,\gamma)=(0.1,-0.2)$, $(0.1,0.2)$, $(0.5,-0.2)$
and $(0.5,0.2)$ respectively. In all those simulations, the mass
resolution is $1.114\times 10^{9}h^{-1}~M_{\bigodot }$; the
particle number is $256^{3}$; the domain grid is a $128\times
128\times 128$ cubic and the finest refined grids have 16384 cells
on each side, corresponding to a force resolution of about
$12h^{-1}~$kpc.

We also run a $\Lambda$CDM simulation with the same physical
parameters and initial condition (see below).

\subsection{Background and Linear Perturbation Evolution}

\label{subsect:bkgd}

Since the coupling between the scalar field and the curvature
produces a time-varying effective gravitational constant, and the
scalar field contributes to the total energy-momentum tensor, we
expect that cosmology in the extended quintessence models is
generally different from $\Lambda$CDM at the background and linear
perturbation levels. A good understanding of this will be helpful
in our analysis of the results from $N$-body simulations, and this
is the subject of this subsection.

Our algorithm and formulae for the background cosmology are detailed in Appendix~\ref{appen:bkgd}, and are
implemented in {\tt MAPLE}. We output the relevant quantities
in a predefined time grid, which could be used (via interpolation)
in the linear perturbation and $N$-body computations.

Fig.~\ref{fig:background} shows the time evolutions of some
background quantities of interests. For ease of
comparison we have chosen $\Omega_m$ and $\Omega_r$ to be the
same in all models including the $\Lambda$CDM one (for definitions
of $\Omega_m$ and $\Omega_r$ see Appendix~\ref{appen:bkgd}), and
as a result in the upper left panel the curves for different
models converge at common righthand ends. We see
increasing $\alpha$ results in an earlier and slower growth of
$\Omega_{\mathrm{DE}}$
($\Omega_{\mathrm{DE}}=1-\Omega_m-\Omega_r$). This indicates a
larger dark energy equation of state parameter, $w$, which is confirmed by
the upper right panel. Physically, this is because, the larger
$\alpha$ is, the steeper the potential becomes and thus the faster
the scalar field rolls. Notice that $w$ is also larger for
positive $\gamma$, with $\alpha$ being the same. This is because
in Eq.~(\ref{eq:sfeom}) the Ricci scalar $R<0$ and for positive
$\gamma$ the term $\frac{R}{2\kappa_{\ast}}f_\varphi$ has the same
sign as $V_\varphi$, thus helping the scalar field to roll faster.
Because of its large predicted value of $w$, the model $(\alpha,
\gamma) = (0.5, 0.2)$ is already excluded by cosmological data,
but here we shall keep it for purely theoretical interest
(\emph{i.e.}, to see how changing $\alpha$ or $\gamma$ changes the
nonlinear structure formation).

We are also interested in how the expansion rate in an extended
quintessence model differs from that in $\Lambda$CDM, and the
results for our models are shown in the lower-left panel of
Fig.~\ref{fig:background}, which plots the
$H/H_{\Lambda\mathrm{CDM}}$ as a function of $a$. The rather odd
behaviour of the models at low redshift is because of the
complicated evolution of the scalar field (and the fact that we
have chosen $H_0$ to be the same for all models, again for ease of
comparison), while the high-redshift behaviour could be seen
directly from Eq.~(\ref{eq:bkgd_friedman}). In
Eq.~(\ref{eq:bkgd_friedman}) the energy density of the scalar
field can be dropped at high $z$, and so we have
\begin{eqnarray}
\left(\frac{\mathcal{H}}{\mathcal{H}_0}\right)^2 &\approx&
\frac{1+f_0}{1+f}\Omega_ma^{-1}
\end{eqnarray}
where we have also neglected the radiation for simplicity (which
is valid after the matter-radiation equality). This shows that in
extended quintessence models the gravitational constant relevant
for the background cosmology is rescaled by $(1+f_0)/(1+f)$.
Because $f_0=f(\varphi_0)$ where $\varphi_0$ is the present-day
value of $\varphi$, and $\varphi$ is monotonically increasing in
time, so for our choice of $f(\varphi)$
[cf.~Eq.~(\ref{eq:coupling_function})] we have $(1+f_0)/(1+f)>1$
for $\gamma>0$ and $(1+f_0)/(1+f)<1$ for $\gamma<0$: thus models
with $\gamma>0$ have $H/H_{\Lambda\mathrm{CDM}}>1$.

It turns out that the gravitational constant relevant for the
growth of matter density perturbations is also different from
the one governing the background cosmology. If we denote the
matter density perturbation by $\delta_m$, then it can be shown,
using the linear perturbation equations, that on small scales the
evolution equation for $\delta_m$ reduces to
\begin{eqnarray}
\delta''_m+\mathcal{H}\delta'_m &=&
G_{\mathrm{N}}\frac{3\mathcal{H}_0^2}{2}\Omega_m\delta_ma^2
\end{eqnarray}
in which $'\equiv d/d\tau$ and $\tau$ is the conformal time (see
Appendix~\ref{appen:bkgd}), and we have defined
\begin{eqnarray}\label{eq:G_N}
G_{\mathrm{N}} &\equiv&
\frac{1+f_0}{1+f}\frac{2+2f+4\left(\frac{df}{d\sqrt{\kappa_\ast}\varphi}\right)^2}
{2+2f+3\left(\frac{df}{d\sqrt{\kappa_\ast}\varphi}\right)^2}.
\end{eqnarray}
Note that this quantity could also be directly read off from the
modified Poisson equation Eq.~(\ref{eq:u_Poisson}).

In the lower right panel of Fig.~\ref{fig:background} we display
the evolution for $G_{\mathrm{N}}$ in the models considered.
Again, $G_{\mathrm{N}}$ is larger at earlier times for positive
$\gamma$ and smaller for negative $\gamma$, because of our
specific choice of $f(\varphi)$ in
Eq.~(\ref{eq:coupling_function}), and the fact that $\varphi$ is
always increasing in time.

It is well known that a higher rate of background expansion means
that structures have less time to form, and a larger
$G_{\mathrm{N}}$ speeds up the structure formation. These two
effects therefore cancel each other to some extent, which results
in a weaker net effect of an extended quintessence field on the
large scale structure formation. This is confirmed by our linear
perturbation computation depicted in Fig.~\ref{fig:linpert}. In
the right-hand panels of this figure we have plotted the matter
power spectra for different models at two different redshifts (0
and 49). It is interesting to note that on small scales the matter
power is closer to that of $\Lambda$CDM, despite the significant
differences in background expansion rate and $G_{\mathrm{N}}$
(cf.~Fig.~\ref{fig:background}). Because of this, we shall choose
$\Lambda$CDM initial condition for our $N$-body simulations for
all our models, saving the effort of generating separate initial
conditions for different models.

The left hand panels of Fig.~\ref{fig:linpert} display the CMB
power spectra for the models we consider. Again the difference
from $\Lambda$CDM is fairly small, and there is only a small shift
of the CMB peaks even though the background expansion rate changes
quite a bit. The latter is because peak positions are determined
by the ratio of the sound horizon size at decoupling and the
angular distance to the decoupling, and in our model both of these
decrease/increase as the Universe expands faster/more slowly,
their ratio does not change much.

To briefly summarise, the study of background cosmology and linear
perturbation shows that a modified background expansion rate and a
rescaled gravitational constant, the two most important factors
affecting structure formation in extended quintessence models are opposite effects. It is then of interest to see how these
two effects compete in the nonlinear regime.

\section{$N$-body Simulation Results}

\label{sect:results}

This section lists the results of extended quintessence $N$-body
simulations. We shall start with a few preliminary results which
both give some basic idea about the extended quintessence effects
and serve as a cross check of our codes. Then we discuss the key
observables for the nonlinear structure formation such as matter
power spectrum, mass function and halo properties. We also comment
on the halo profile of the scalar field and the spatial variation
of gravitational constant.

\subsection{Preliminary Results}

As mentioned above, in both the linear and $N$-body codes we
compute background quantities via an interpolation of some
pre-computed table. Because background cosmology is important in
determining the structure formation, it is important to check its
accuracy. For this we have recorded in Table~\ref{tab:background}
The age of the universe today for different models as computed by
these two codes. The two codes are compatible with each other
indeed.
\begin{table}[htbp]
\caption{\label{tab:background} Current age of the Universe for
the different models under consideration as computed by the linear
perturbation and $N$-body codes. Unit is Gyr.}
\end{table}
\begin{center}
\begin{tabular}{cccccccccc}
\hline\hline
model & linear code & $N$-body code\\
\hline
$\Lambda$CDM & 13.680 & 13.678 \\
$(\alpha,\gamma)=(0.1,-0.2)$ & 13.639 & 13.638 \\
$(\alpha,\gamma)=(0.1,0.2)$ & 13.408 & 13.408 \\
$(\alpha,\gamma)=(0.5,-0.2)$ & 13.513 & 13.513 \\
$(\alpha,\gamma)=(0.5,0.2)$ & 12.097 & 12.096 \\
\hline \hline
\end{tabular}
\end{center}

Because one of the advantages of our $N$-body code is that it
solves the scalar field perturbation explicitly, it is important
to check that the solution is with expectations. From
Eqs.~(\ref{eq:WFphiEOM2}, \ref{eq:WFPoisson2}) it could be seen
clearly that, if the contribution to the local density and
pressure from the scalar field is negligible compared with that
from matter, then the modified Poisson equation and scalar field
equation of motion end up with the same source term (up to a
$\bar{\varphi}$-dependent coefficient). In this situation we
expect
\begin{eqnarray}\label{eq:aid3}
u &=& -\frac{2\gamma\sqrt{\kappa_{\ast}}\bar{\varphi}}
{1+\frac{8\gamma^2\kappa_\ast\bar{\varphi}^2}{1+\gamma\kappa_\ast\bar{\varphi}^2}}\Phi_c,
\end{eqnarray}
which means that $u$ is simply proportional to $\Phi_c$ with a
time-dependent coefficient. In Fig.~\ref{fig:scal_pot} we have
checked this relation explicitly: we select a thin slice of
the simulation box, fetch the values for $u$ and $\Phi_c$ at the
positions of the particles (about 10000 in total) therein, and
display them as scatter plots. The solid curve is the
approximation Eq.~(\ref{eq:aid3}) while the green dots are
simulation results; we can see they agree very well with each
other, showing that the above approximation is a good one.
Note that the scalar field perturbation
$a\sqrt{\kappa_{\ast}}\delta\varphi$ is generally less than
$10^{-6}$, compared with the background value
$\sqrt{\kappa_{\ast}}\bar{\varphi}\sim\mathcal{O}(0.1-1)$. This
confirms that it is consistent to neglect the perturbation in
scalar field density/pressure, drop terms such as
$\dot{\delta\varphi}$ and $\ddot{\delta\varphi}$, and replace
$\varphi$ by $\bar{\varphi}$ in coefficients of perturbation
quantities such as
$\vec{\partial}^{2}_{\mathbf{x}}\left(a\sqrt{\kappa_{\ast}}\delta\varphi\right)$
and $\vec{\partial}^{2}_{\mathbf{x}}\Phi$. It also serves as a
check of the numerical code.

As a final consistency check, let us consider the total
gravitational force on particles. In extended quintessence models,
this is given by Eq.~(\ref{eq:WFPoisson2}), and when the
perturbation in the scalar field density/pressure is negligible
(which is the case as shown above) we get
\begin{eqnarray}
\nabla^{2}\Phi_c &\approx&
\frac{3}{2}G_{\mathrm{N}}\Omega_mH_0^2\left(\rho_c-1\right)
\end{eqnarray}
in which $G_{\mathrm{N}}$ is given in Eq.~(\ref{eq:G_N}). On the
other hand, if we consider (na\"{\i}vely) that gravity is
described by general relativity, then we should neglect the $G_{\mathrm{N}}$ on the
right-hand side. Manipulating Eqs.~(\ref{eq:WFphiEOM2},
\ref{eq:WFPoisson2}) we obtain:
\begin{eqnarray}
\frac{1+\gamma\kappa_{\ast}\bar{\varphi}_0^2}{1+\gamma\kappa_{\ast}\bar{\varphi}^2}\nabla^{2}
\left(\Phi_c+\frac{\gamma\sqrt{\kappa_{\ast}}\bar{\varphi}}{1+\gamma\kappa_{\ast}\bar{\varphi}^2}u\right)
&\approx& \frac{3}{2}\Omega_mH_0^2\left(\rho_c-1\right).\nonumber
\end{eqnarray}
Thus
$\frac{1+\gamma\kappa_{\ast}\bar{\varphi}_0^2}{1+\gamma\kappa_{\ast}\bar{\varphi}^2}
\left(\Phi_c+\frac{\gamma\sqrt{\kappa_{\ast}}\bar{\varphi}}{1+\gamma\kappa_{\ast}\bar{\varphi}^2}u\right)$
acts as the potential for na\"{\i}ve gravity (\emph{i.e.},
general relativity), and by differcing it we could obtain the na\"{\i}ve
gravitational force. In Fig.~\ref{fig:force} we show the scatter
plot of the na\"{\i}ve gravity versus full gravity for the same
particles as in Fig.~\ref{fig:scal_pot} (green dots) as well as
their approximate ratio $G_{\mathrm{N}}$ (solid line). Again, the
agreement is remarkably good.

\begin{figure}[tbp]
\centering \includegraphics[scale=0.5] {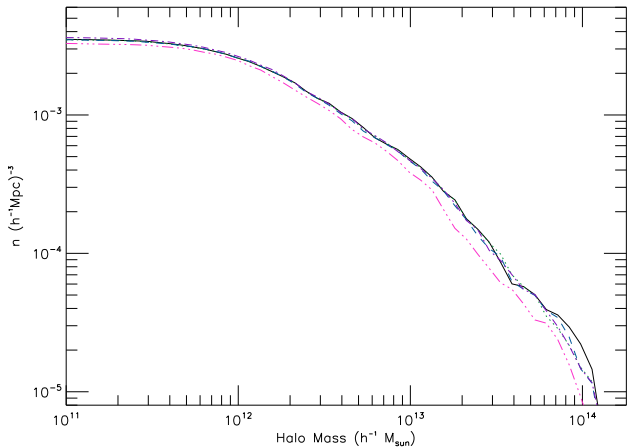} \caption{(Color
Online) The mass functions for the models considered. The black
solid, green-dotted, blue-dashed, purple dot-dashed and pink
dot-dot-dot-dashed curves stand for the results for $\Lambda$CDM
and extended quintessence models with
$(\alpha,\gamma)=(0.1,-0.2)$, $(0.1,0.2)$, $(0.5,-0.2)$ and
$(0.5,0.2)$ respectively. The horizontal axis denotes the halo
mass (in unit of $h^{-1}M_{\bigodot}$) and the vertical axis is
the halo number density (in unit of $h^{3}\mathrm{Mpc}^{-3}$).
Only the results at $a=1$ are plotted.} \label{fig:mf}
\end{figure}

\begin{figure*}[tbp]
\centering \includegraphics[scale=1.0] {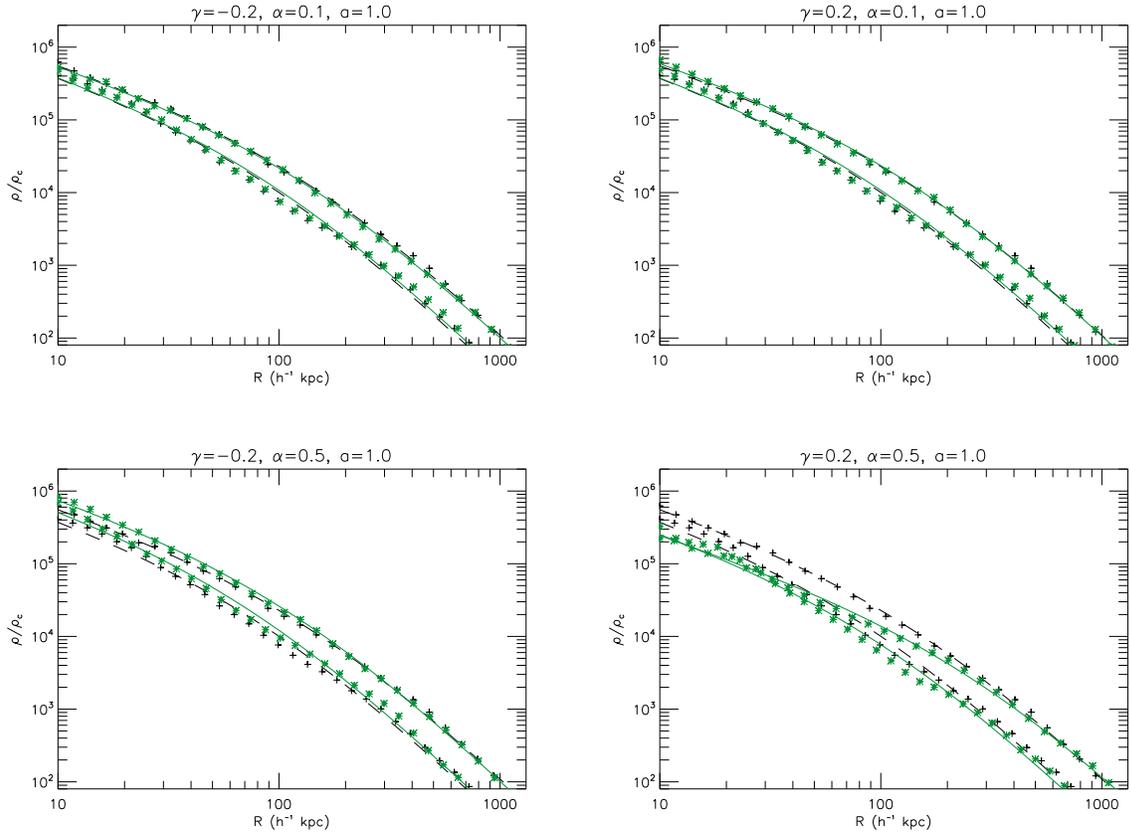}
\caption{(Color Online) The NFW fitting results for two halos
randomly selected from the 80 most massive halos in each
simulation (see text for details). The upper and lower green
asterisks represent respectively the density profile from $N$-body
simulation for the more and less massive halo, and the green solid
curves their NFW fittings. For comparison we also shown the
corresponding $N$-body (black crosses) and fitting (black dashed
curves) results for the $\Lambda$CDM model. The horizontal axis is
the distance from halo centre (in units of $h^{-1}$kpc) and the
vertical axis is the density contrast. The four panels are for the
four models as indicated above the frames.} \label{fig:NFW_dens}
\end{figure*}

\begin{figure*}[tbp]
\centering \includegraphics[scale=1.0] {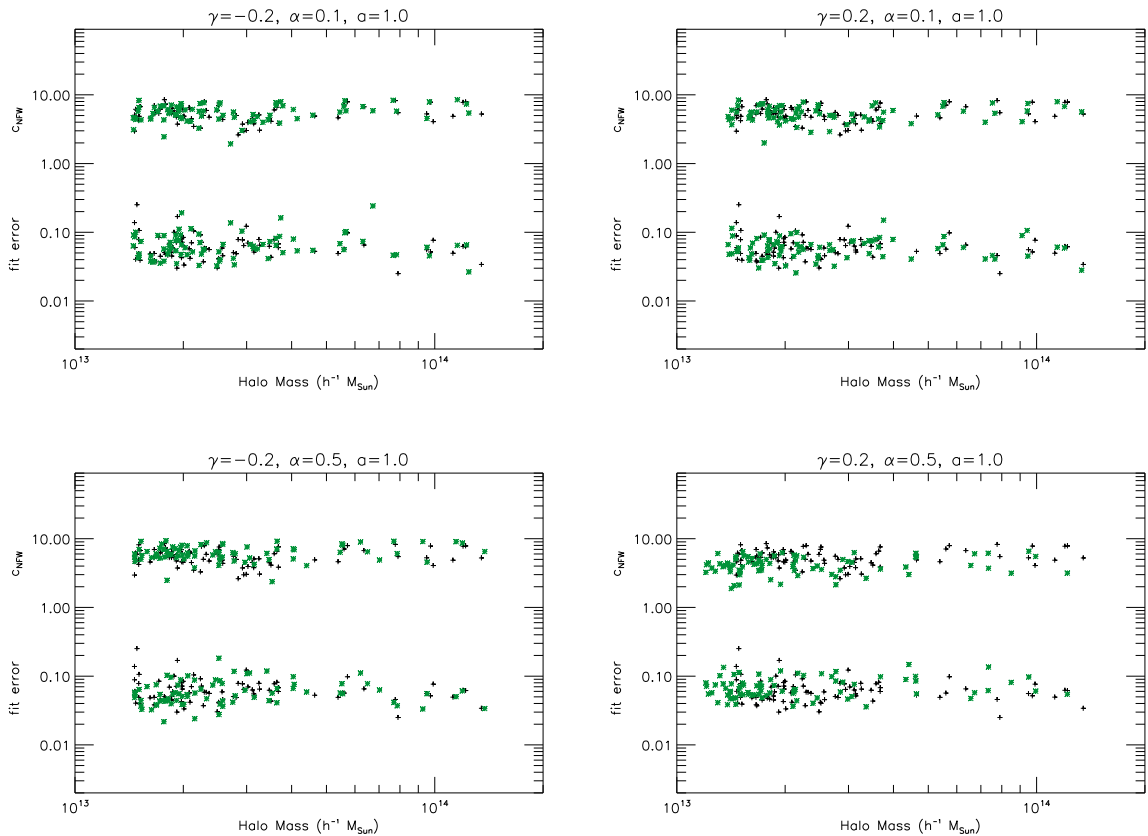}
\caption{(Color Online) Scatter plot of the NFW fitting of the
dark matter halo density profiles for the 80 most massive halos in
each simulation box. In all panels the black crosses and green
asterisks stand for results of $\Lambda$CDM and extended
quintessence models respectively. The upper cluster of points in
each panel represents the fitted $c_{\mathrm{NFW}}$ and the lower
cluster is for the fitting error, as indicated beside the
vertical axis. The horizonal axis is the halo mass (in units of
$h^{-1}M_{\bigodot}$). The four panels are for the models as
indicated above the frames.} \label{fig:NFW_fit}
\end{figure*}

\begin{figure*}[tbp]
\centering \includegraphics[scale=0.95] {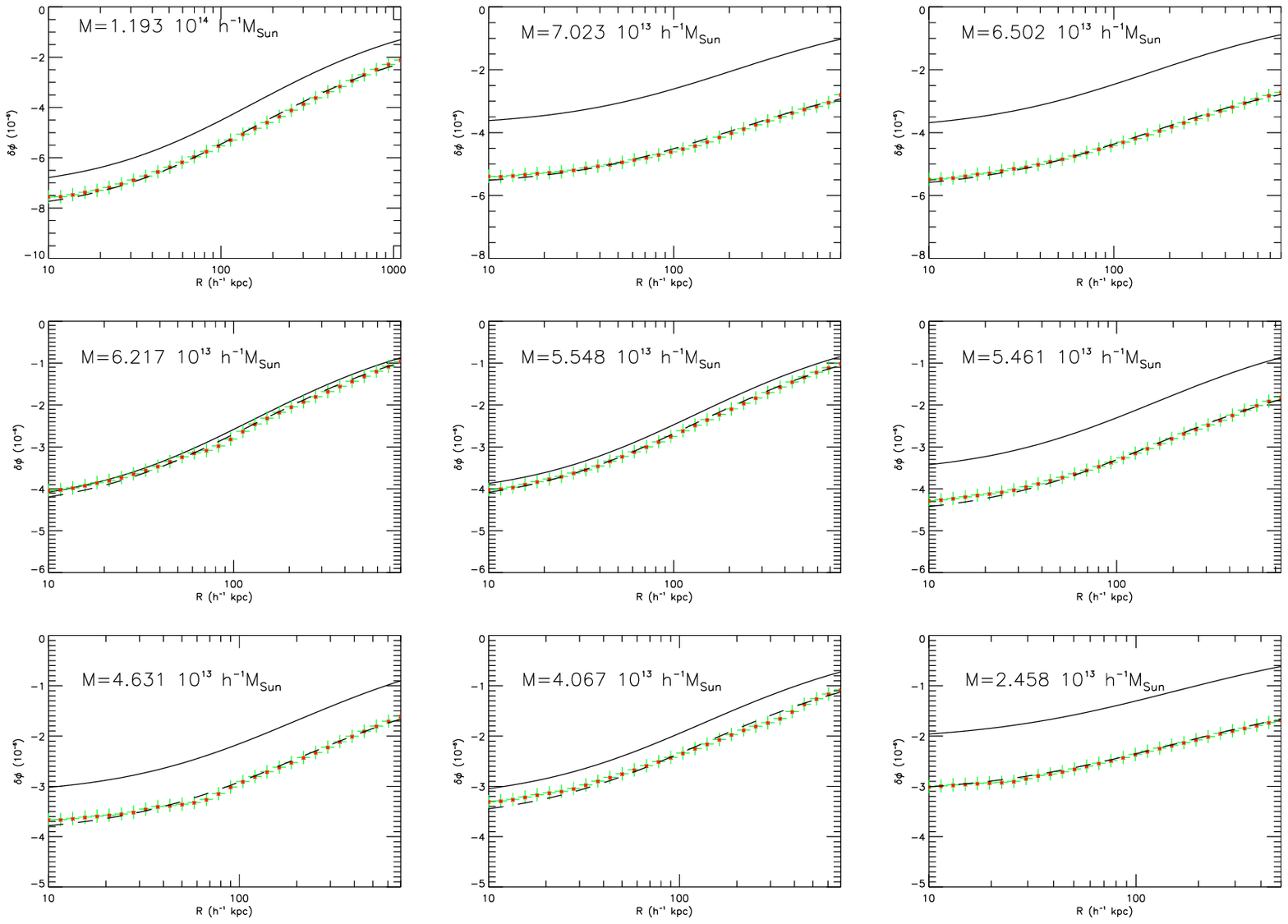}
\caption{(Color Online) Comparison between our analytic formula
for the scalar field perturbation
$a\sqrt{\kappa_{\ast}}\delta\varphi$ [Eq.~(\ref{eq:delta_varphi})]
and the results from numerical simulation. The nine panels are for
nine halos selected from the simulation box, whose masses are
indicated inside each frame. In each panel the solid curve is
Eq.~(\ref{eq:delta_varphi}) with $\Phi_\ast=0$, green crosses are
the numerical results for $a\sqrt{\kappa_{\ast}}\delta\varphi$,
the dashed curve is Eq.~(\ref{eq:delta_varphi}) with $\Phi_\ast$
appropriately tuned to match the green crosses, and the red
asterisks are the $a\sqrt{\kappa_{\ast}}\delta\varphi$ computed
from the value of $\Phi$ using Eq.~(\ref{eq:aid3}). The horizontal
axis is the distance from the halo centre, and vertical axis
stands for the value of $a\sqrt{\kappa_{\ast}}\delta\varphi$.}
\label{fig:NFW_scal}
\end{figure*}

\subsection{Nonlinear Matter Power Spectrum}

As we have seen above, the linear matter power spectrum for the
extended quintessence model really does not show much useful
information on small scales, and so we need to investigate whether
nonlinear effects could change this situation and therefore
potentially place more meaningful constraints.

Fig.~\ref{fig:power} provides a positive answer to this
question. Here we have plotted the fractional difference of the
extended quintessential nonlinear matter power spectrum from that
for $\Lambda$CDM (remember that we use the same initial condition
for all simulations). We can see that for the models with
$\alpha=0.1$ the differences are small even in the nonlinear
regime, indicating that the scalar field really does not affect
the matter distribution significantly if the potential is flat.
However, for the $\alpha=0.5$ cases in which the coupling strength
$\gamma$ remains the same, the difference could be as large as
$30\%\sim50\%$, guaranteeing an observable signature.

Furthermore, for negative $\gamma$ (the purple curve) the extended
quintessential power spectrum beats the $\Lambda$CDM one on small
scales, whereas for the positive $\gamma$ case (the pink curve) it
is just the opposite. As shown before, when $\gamma<0$, both the
background expansion rate and the effective gravitational constant
governing the structure formation decrease, boosting and weakening
the collapse of matter respectively. In our
$\alpha=0.5$ cases the first effect has clearly taken over on small scales.

\subsection{Mass Function}

A second important observable is the mass
function. This gives the number density of dark matter halos as a
function of halo mass. For this we need to identify the dark
matter halos from the output particle distribution of the $N$-body
simulations, and this determination is performed using a modified version of
{\tt MHF} \citep{mhf}, {\tt MLAPM}'s default halo finder.

\texttt{MHF} optimally utilizes the refinement structure of the
simulation grids to pin down the regions in which potential halos
reside and organize the refinement hierarchy into a tree
structure. \texttt{MLAPM} refines grids according to the particle
density on them and so the boundaries of the refinements are
simply isodensity contours. \texttt{MHF} collects the particles
within these isodensity contours (as well as some particles
outside). It then performs the following operations: (i) assuming
spherical symmetry of the halo, calculate the escape velocity
$v_{esc}$ at the position of each particle, (ii) if the velocity
of the particle exceeds $v_{esc}$ then it does not belong to the
virialized halo and is removed. Steps (i) and (ii) are then
iterated until all unbound particles are removed from the halo or
the number of particles in the halo falls below a pre-defined
threshold, which is 20 in our simulations. Note that the removal
of unbound particles is not used in some halo finders using the
spherical overdensity (SO) algorithm, which includes the particles
in the halo as long as they are within the radius of a virial
density contrast. Another advantage of \texttt{MHF} is that it
does not require a predefined linking length in finding halos,
such as the friend-of-friend procedure.

Our modification to {\tt MHF} is simple: because the effective
gravitational constant in the extended quintessence models is
rescaled by a factor $G_{\mathrm{N}}$ [cf.~Eq.~(\ref{eq:G_N})],
the escape velocity of particles from a halo is also multiplied by
this factor, and in {\tt MHF} we have only changed the criterion
for removing particles from virialised halos accordingly. In
reality, because we are only interested in the $a=1$ halos in this
work, $G_{\mathrm{N}}$ is quite close to 1 and the effect of our
modification is not large.

The mass functions for our simulated models are shown in
Fig.~\ref{fig:mf}. It shows that all extended quintessence models
considered here, irrespective of their parameters, produce less
massive halos than $\Lambda$CDM, whereas (only) the model
$(\alpha,\gamma)=(0.5,-0.2)$ produces a larger number of less
massive halos. These features are in broad agreement with those
shown in the matter power spectra (Fig.~\ref{fig:power}) where all
models show less matter clustering on the large scales, whereas
(only) the model $(\alpha,\gamma)=(0.5,-0.2)$ shows more power on
small scales. The physical reason is again the competition between
the modified background expansion rate and rescaled effective
gravitational constant $G_{\mathrm{N}}$.

\subsection{Halo Properties}

In the $\Lambda$CDM paradigm, it is well known that the internal
density profiles of dark matter halos are very well described by
the Navarro-Frenk-White \citep{NFW} formalism
\begin{eqnarray}\label{eq:NFW}
\frac{\rho(r)}{\rho_{c}} &=&
\frac{\beta}{\frac{r}{R_s}\left(1+\frac{r}{R_s}\right)^2}
\end{eqnarray}
where $\rho_c$ is the critical density for matter, $\beta$ is a
dimensionless fitting parameter and $R_s$ a second fitting
parameter with length dimension. $\beta$ and $R_s$ are generally
different for different halos and should be fitted for individual
halos, but the formula Eq.~(\ref{eq:NFW}) is quite universal.

We are thus interested in whether the halo profiles in an extended
quintessential Universe are also featured by this universal form.
For this we select the 80 most massive halos from each simulation
and fit their density profiles to Eq.~(\ref{eq:NFW}). The results
show that the NFW profile describes the extended quintessential
halos at least as well as it does for the $\Lambda$CDM halos.
Fig.~\ref{fig:NFW_dens} shows the fittings for two halos randomly
picked out of the 80: one at $\sim(10.34, 28.63, 13.91)h^{-1}$Mpc
with mass $\sim1.88\times10^{14}M_{\bigodot}$ and the other at
$(41.77,31.91,21.20)h^{-1}$Mpc with a mass
$\sim4.98\times10^{13}M_{\bigodot}$.

There are some interesting features in Fig.~\ref{fig:NFW_dens}.
Firstly, for the models with $\alpha=0.1$ (the top panels) the
halo density profile for extended quintessence models (green
asterisks) is very similar to the $\Lambda$CDM results (black
crosses) and thus their fittings almost coincide. Secondly, for
the model of $(\alpha,\gamma)=(0.5,-0.2)$, the chosen halos show
more concentration of the density profiles in the scalar model
than in $\Lambda$CDM. Thirdly, the model of
$(\alpha,\gamma)=(0.5,-0.2)$ has just the opposite trend and
suffers a suppression of density in large parts of chosen halos.

To verify that the above features are actually typical for the
corresponding models, we have plotted in Fig.~\ref{fig:NFW_fit}
the fitting results for all the 80 massive halos in all simulated
models. Here in addition to the NFW concentration parameter
$c_{\mathrm{NFW}}=r_{200}/R_s$, where $r_{200}$ is the radius at
which the density is equal to 200 times the critical density
$\rho_c$ and $R_s$ the NFW parameter, we have also shown the fitting
errors for each halo.

We would like to point out several important implications of
Fig.~\ref{fig:NFW_fit}. Firstly, for all models the fitting error
for the extended quintessential halos (lower green asterisks) is
comparable to that for the $\Lambda$CDM halos (lower black
crosses), indicating that the density profiles for the former are
equally well described by the NFW formula Eq.~(\ref{eq:NFW}).
Secondly, for the models with $\alpha=0.1$ (the top panels) we can
see that the fitted $c_{\mathrm{NFW}}$ for the extended
quintessential halos is comparable to that for $\Lambda$CDM, which
is in agreement with our finding in Fig.~\ref{fig:NFW_dens} that
the density profiles for the chosen halos are almost the same as in
the $\Lambda$CDM prediction. Thirdly, for the model of
$(\alpha,\gamma)=(0.5,-0.2)$, the halos tend to be \emph{more}
concentrated (\emph{i.e.}, with \emph{larger} $c_{\mathrm{NFW}}$)
than in $\Lambda$CDM. Fourthly, for the model
$(\alpha,\gamma)=(0.5,0.2)$, the halos tend to be \emph{less}
concentrated (\emph{i.e.}, with \emph{smaller} $c_{\mathrm{NFW}}$)
than in $\Lambda$CDM. The above three features show that our
qualitative findings in Fig.~\ref{fig:NFW_dens} are quite typical.
Finally, the halo masses in the model of
$(\alpha,\gamma)=(0.5,-0.2)$ are on average smaller than those in
$\Lambda$CDM, because the upper green asterisks in the lower right
panel consistently shift leftwards with respect to the upper black
crosses: this is consistent with the mass function result that
this model produces less massive halos than $\Lambda$CDM.

In summary, the halo density profiles for the extended 
quintessence models are well described by the NFW formula, but the existence
of the scalar field and in particular its coupling to curvature do
change the concentration parameters of the halos, so long as the potential
is not too flat. It seems that the modified background expansion rate
beats the effect of the rescaled effective gravitational constant here.

\subsection{Halo Profile for Scalar Field Perturbation}

We have already seen that the coupling between the scalar field
and the curvature scalar causes time and spatial variations of the
locally measured gravitational constant $\kappa_{\bigoplus}$. It
is then of our interest to ask how $\kappa_{\bigoplus}$ varies
across a given halo and whether this could produce observable
effects. This subsection answers this question, by
giving an analytical formula and comparing it with numerical
results.

Recall that Fig.~\ref{fig:scal_pot} shows that to a high
precision the scalar field perturbation
$a\sqrt{\kappa_{\ast}}\delta\varphi$ is proportional to the
gravitational potential $\Phi$ [cf.~Eq.~(\ref{eq:aid3})]
everywhere. This means that if we could derive an analytical
formula for $\Phi$ in halos, then we know
$a\sqrt{\kappa_{\ast}}\delta\varphi$ straightforwardly. Such a
derivation has been done in \citep{Li:2010alpha} for a different model,
but here we shall briefly repeat it for the extended quintessence
model for completeness.

Assuming Eq.~(\ref{eq:NFW}) as the density profile and sphericity
of halos, we can derive $V_c(r)$, the circular velocity of a
particle moving around the halo at a distance $r$ from halo
centre, to be
\begin{eqnarray}\label{eq:V_c}
V_c^2(r) &=& \frac{GM(r)}{r}\nonumber\\
&=& 4\pi
G\beta\rho_cR^3_s\left[\frac{1}{r}\ln\left(1+\frac{r}{R_s}\right)
- \frac{1}{R_s+r}\right]\ \
\end{eqnarray}
where $M(r)$ is the mass enclosed in radius $r$, $G$ is the
\emph{properly rescaled} gravitational constant. Again, this
equation is parameterized by $\beta$ and $R_s$. From a simulation
point of view, it is straightforward to measure $M(r)$ and then
use Eq.~(\ref{eq:V_c}), instead of Eq.~(\ref{eq:NFW}), to fit the
values of $\beta$ and $R_s$; from an observational viewpoint, it
is easy to measure $V_c(r)$, which could again be used to fit
$\beta$ and $R_s$.

The potential inside a spherical halo is then given as
\begin{eqnarray}
\Phi(r) &=& \int^{r}_{0}\frac{GM(r')}{r'^2}dr' + C
\end{eqnarray}
in which $GM(r)/r^2$ is the gravitational force and $C$ is a
constant to be fixed using the fact that
$\Phi(r=\infty)=\Phi_\infty$ where $\Phi_\infty$ is the value of
the potential far from the halo.

Using the formula for $GM(r)/r^2$ given in Eq.~(\ref{eq:V_c}) it
is not difficult to find that
\begin{eqnarray}
\int^{r}_{0}\frac{GM(r')}{r'^2}dr' &=& 4\pi
G\beta\rho_cR^3_s\left[\frac{1}{R_s}-\frac{\ln\left(1+\frac{r}{R_s}\right)}{r}\right]\nonumber
\end{eqnarray}
and so
\begin{eqnarray}
C &=& \Phi_\infty - 4\pi G\beta\rho_cR^2_s.
\end{eqnarray}
Then it follows that
\begin{eqnarray}\label{eq:Phi}
\Phi(r) &=& \Phi_\infty - 4\pi
G\beta\rho_c\frac{R^3_s}{r}\ln\left(1+\frac{r}{R_s}\right).
\end{eqnarray}
If the halo is \emph{isolated}, then $\Phi_\infty=0$ and we get
\begin{eqnarray}\label{eq:Phi_isolated_halo}
\Phi(r) &=& - 4\pi
G\beta\rho_c\frac{R^3_s}{r}\ln\left(1+\frac{r}{R_s}\right).
\end{eqnarray}
However, in $N$-body simulations, we have a large number of dark
matter halos and no halo is totally isolated from the others. In
such situations, $\Phi_\infty$ in Eq.~(\ref{eq:Phi}) should be
replaced by $\Phi_\ast$, which is the potential produced by other
halos inside the considered halo (note that in practice
$\Phi_\ast$ could be position dependent as well, but for
simplicity we assume that it is a constant, which is a good
assumption for many halos). Then we get
\begin{eqnarray}\label{eq:delta_varphi}
a\sqrt{\kappa}\delta\varphi(r) &=& -\frac{2\gamma\sqrt{\kappa_{\ast}}\bar{\varphi}}
{1+\frac{8\gamma^2\kappa_{\ast}\bar{\varphi}}{1+\gamma\kappa_\ast\bar{\varphi}}}\nonumber\\
&&\times\left[\Phi_\ast - 4\pi
G\beta\rho_c\frac{R^3_s}{r}\ln\left(1+\frac{r}{R_s}\right)\right].\
\ \
\end{eqnarray}

Eq.~(\ref{eq:delta_varphi}) provides a neat analytical formula for
$a\sqrt{\kappa}\delta\varphi$ in halos, but unfortunately in most
cases it cannot be used directly because we lack information about
$\Phi_\ast$. We will then be forced either to fit $\Phi_\ast$ as a
free parameter, or tune its value to match simulations or
observations. In this work we shall take the second approach, and
we find that with an appropriate value of $\Phi_\ast$ and
with values of $\beta$ and $R_s$ fitted using Eq.~(\ref{eq:V_c}),
Eq.~(\ref{eq:delta_varphi}) agrees with numerical results for most
halos.

Some examples are shown in Fig.~\ref{fig:NFW_scal}, in which we have
computed $a\sqrt{\kappa}\delta\varphi(r)$ using four different methods:
direct $N$-body simulation results (big green crosses), Eq.~(\ref{eq:delta_varphi}) with $\Phi_{\ast }=0$ (solid curves), 
Eq.~(\ref{eq:delta_varphi}) with $\Phi_{\ast}$ properly tuned (dashed curves) and
Eq.~(\ref{eq:aid3}) with $\Phi$ directly from $N$-body simulations (small
red asterisks). Clearly the crosses and asterisks agree with each
other very well, which is another demonstration that Eq.~(\ref{eq:aid3}) is a very good approximation (cf.~Fig.~\ref{fig:scal_pot}). The
solid curves different significantly from the numerical results, showing
that $\Phi_{\ast}$ is actually nonzero; once it is appropriately tuned,
then Eq.~(\ref{eq:delta_varphi}) (dashed curves) agree with the numerical
results very well for all the chosen halos. Eq.~(\ref{eq:delta_varphi}) 
therefore provides a useful analytical formula which might aid
in general analysis.

We also notice that across the halos, the variation of $a\sqrt{\kappa}\delta\varphi$ is typically $\lesssim\mathcal{O}\left(10^{-6}\sim10^{-5}\right)$. 
Such a small variation is unlikely to be detectable using current
observational instruments, and thus we do not expect special constraints
based on the spatial variation of $G$. However, we stress that the above
result is only for a class of extended quintessence models, and although we
expect it to be valid for other potentials which are not particularly
nonlinear, the situation could be dramatically changed in cases where the
potential or coupling function becomes highly nonlinear. Such models require
a more careful treatment, including some of the approximations adopted above
becoming invalid, and are thus beyond the scope of the current work.

\section{Summary and Conclusion}

\label{sect:con}

In summary, in this paper we have described a numerical method
to study extended quintessence models, where the quintessence field
has a scalar-tensor type of coupling to the curvature, from background
cosmology to nonlinear structure formation, and discussed the regime of
validity of the method. Instead of assuming a Yukawa force due to scalar
coupling or simply a rescaling of gravitational constant, we have solved the
scalar field and its spatial variation explicitly from their equation of
motion. This is a necessary step in general to obtain trustable results and
check various approximations which are made to simplify the computation.

As specific examples, we apply the above method to a specific class of
models with inverse power-law potential Eq.~(\ref{eq:potential}) and
non-minimal coupling Eq.~(\ref{eq:coupling_function}). The analysis of 
the background cosmology and its linear perturbation shows that for
these models the effective gravitational 'constants' relevant for the
cosmic expansion rate and structure formation are either both increased or
both decreased (albeit by slightly different amounts). The two effects
compete and cancel each other, and as a result the net effect on
large scale structure in the linear regime is weak (cf.~Fig.~\ref{fig:linpert}). We then investigated whether a more
significant signature of the scalar field could be imprinted in the
nonlinear regime of structure formation.

The nonlinear matter power spectra plotted in Fig.~\ref{fig:power} suggests that the effect of the scalar field is more significant in the
nonlinear regime. For the models with $\alpha =0.5$ (\emph{i.e.}, steeper
potential), the scalar field changes (either increases or decreases) the
matter power spectrum by $30\sim 50\%$ on small scales with respect to the $\Lambda$CDM prediction. Going to nonlinear scales thus greatly
enhances the power of constraining such models using cosmological data.
However, the power is more limited for models with $\alpha=0.1$ (\emph{i.e.}, shallower potential); their matter power spectra are very similar
to the $\Lambda$CDM results.  Of the two competing effects
mentioned above, we find that the modified background expansion rate is
more influential on nonlinear scales.

Properties of mass functions (cf.~Fig.~\ref{fig:mf}) are in qualitative
agreement with what we have seen in the matter power spectrum, with the
extended quintessence models producing less massive halos than $\Lambda$CDM.
Therefore galaxy cluster counts could place meaningful constraints on such
models as well. But as the matter power spectrum, the mass function for the
models with $\alpha=0.1$ (\emph{i.e.}, shallower potential) is very similar
to the $\Lambda$CDM result.

The halo density profiles for the extended quintessence models are
shown to be well described by the well-known NFW formula (cf.~Fig.~\ref{fig:NFW_dens}, \ref{fig:NFW_fit}). In Fig.~\ref{fig:NFW_fit} we have shown
the results of the fitting for the 80 most massive halos from each
simulation. Consistent with the findings in Figs.~\ref{fig:power} and 
\ref{fig:mf}, we see that the concentration parameter $c_{\mathrm{NFW}}$ for
the halos in the $\alpha =0.1$ models is almost the same as for the $\Lambda 
$CDM halos. But for $\alpha =0.5$, the $\gamma =-0.2$ and $\gamma =0.2$
cases predict overall bigger and smaller $c_{\mathrm{NFW}}$ than $\Lambda $CDM, respectively. Furthermore, Fig.~\ref{fig:NFW_fit} shows
clearly that the halos in the $(\alpha ,\gamma )=(0.5,0.2)$ model are
consistently less massive than those in $\Lambda $CDM, as suggested by the
mass function plots.

Scalar-tensor theories (which the extended quintessence models belong
to) are often studied in the context of varying gravitational constant, 
and so we have also considered the spatial variations (time variation
has been investigated in detail elsewhere and will not be repeated
here) in the scalar field (or equivalently the locally measured
gravitational constant $\kappa_{\bigoplus }$). We first showed in
Fig.~\ref{fig:scal_pot} that the approximation that the scalar field
perturbation $a\sqrt{\kappa_{\ast }}\delta\varphi$ is proportional to the
gravitational potential $\Phi$ [cf.~Eq.~(\ref{eq:aid3})] is fairly
accurate. Then, based on this fact and using the NFW density profile, we
derive an analytical formula for $a\sqrt{\kappa_{\ast }}\delta \varphi(r)$
in spherical halos, in which the parameters are obtained by fitting the NFW
circular velocity profile. We have shown that this formula could be tuned to
fit the numerical results pretty well for most halos (cf.~Fig.~\ref{fig:NFW_scal}).

Fig.~\ref{fig:NFW_scal} indicates that the spatial variation of $a\sqrt{\kappa_{\ast}}\delta\varphi$ across halos is at most of order $10^{-5}$,
which is far smaller than the background value $\sqrt{\kappa_{\ast}}\bar{\varphi}\sim\mathcal{O}(0.1-1)$. Therefore the spatial variation of 
$\kappa_{\bigoplus}$ is expected to be of order $10^{-5}$ or less in the
halos, which is difficult to detect.

The smallness of $a\sqrt{\kappa_{\ast }}\delta \varphi$ also implies that
the approximations we have made to simplify the simulations are valid. For 
example, because $|a\sqrt{\kappa_{\ast }}\delta \varphi |\ll 1$, 
which means it is reasonable to ignore the contribution from $\dot{\delta\varphi}, \ddot{\delta\varphi}$ to the total density/pressure 
perturbation, we can also replace $\varphi $ by $\bar{\varphi}$ in the
coefficients of perturbation quantities such as $\vec{\partial}_{\mathbf{x}}^{2}\Phi$ and $\vec{\partial}_{\mathbf{x}}^{2}\left( a\sqrt{\kappa_{\ast }}\delta \varphi \right) $.
Moreover, the quasi static limit, \emph{i.e.}, neglecting $\dot{\delta\varphi }, \ddot{\delta\varphi}$ compared to 
$\vec{\partial}_{\mathbf{x}}^{2}\left(a\sqrt{\kappa_{\ast}}\delta\varphi\right)$, is guaranteed to work well.

One of the most important results of this work is that it confirms
explicitly that, for a broad range of extended quintessence models, the $N$-body simulation reduces to modifying the background expansion rate
and rescaling the effective gravitational constant based on the the
background value of $\varphi$. This works to quite high accuracy and thus
there is no need to solve the scalar field equation of motion explicitly,
which is particularly time-consuming for large simulations. However,
we expect this approximation to break down in extreme situations 
where the potential (or perhaps the coupling function) becomes
highly nonlinear, and then both our results and method might have to be
revised.

\begin{acknowledgments}
The work described in this paper has been performed on
\texttt{TITAN}, the computing facilities at the University of Oslo
in Norway; coding and testing are done on the {\tt SARA}
supercomputer in the Netherlands, supported by the European
Community Research Infrastructure Action under the FP8
"Structuring the European Research Area" Programme. Postprocessing
is done on \texttt{COSMOS}, the UK's National Cosmology
Supercomputer. We have used {\tt POWMES} \citep{powmes} to measure
the matter power spectrum from output particle distribution, and a
modified version of {\tt CAMB} \citep{Lewis:2000} for our linear
perturbation computation. We thank David Wands for discussions. B.~Li is supported by the Research
Fellowship at Queens' College, Cambridge, and the Science and
Technology Facility Council of the United Kingdom.   DFM thanks the Research Council of Norway FRINAT grant 197251/V30.
\end{acknowledgments}

\bigskip

\appendix

\section{Useful Expressions}

\label{appen:expression}

In this appendix we list some useful expressions in the derivation
of our equations, because different researchers use different
conventions.

Our line element is
\begin{eqnarray}
ds^2 &=& a^2(1+2\phi)d\tau^2 - a^2(1-2\psi)\gamma_{ij}dx^idx^j
\end{eqnarray}
where $\tau$ is the conformal time, $x^i$ is the comoving
coordinate and $\gamma_{ij}$ the metric in the 3-space (with $i,
j$ running over $1,2,3$). The nonzero Christofle symbols, up
to first order in perturbation, are
\begin{eqnarray}
\Gamma^{0}_{00} &=& \frac{a'}{a}+\phi',\ \ \ \ \ \ \ \ \ \ \ \ \ \
\Gamma^{0}_{0i}\ =\
\phi_{,i}\nonumber\\
\Gamma^{i}_{0j} &=& \left(\frac{a'}{a}-\psi'\right)\delta^{i}_{\
j}, \ \ \ \ \ \Gamma^{i}_{00}\ =\ \phi^{,i}\nonumber\\
\Gamma^{0}_{ij} &=&
\frac{a'}{a}(1-2\phi-2\psi)\gamma_{ij}-\psi'\gamma_{ij}\nonumber\\
\Gamma^{i}_{jk} &=& -\psi_{,k}\delta^{i}_{\ j} -
\psi_{,j}\delta^{i}_{\ k}+\psi^{,i}\gamma_{jk}
\end{eqnarray}
where a comma denotes a partial derivative with respect to the
comoving coordinate, and indices are raised and lowered by $\gamma^{ij}$ and $\gamma_{ij}$ respectively. $^{\prime}\equiv d/d\tau$.

The Ricci tensor is 
\begin{eqnarray}
R_{ab} &=& \Gamma^{c}_{ab,c} - \Gamma^{c}_{ac,b} +
\Gamma^{c}_{cd}\Gamma^{d}_{ab} - \Gamma^{d}_{cb}\Gamma^{c}_{ad}
\end{eqnarray}
and its components up to first order in perturbation are
\begin{eqnarray}
R_{00} &=& \phi^{,i}_{\ ,i} -
3\left[\frac{a''}{a}-\left(\frac{a'}{a}\right)^2\right] + 3\psi''
+ 3\frac{a'}{a}\left(\phi'+\psi'\right),\nonumber\\
R_{0i} &=& 2\psi'_{,i} + 2\frac{a'}{a}\phi_{,i},\nonumber\\
R_{ij} &=& -\psi''\gamma_{ij} -
\frac{a'}{a}\left(\phi'+5\psi'\right)\gamma_{ij} -
(\phi-\psi)_{,ij}\nonumber\\
&&+\left[\frac{a''}{a}+\left(\frac{a'}{a}\right)^2\right](1-2\phi-2\psi)\gamma_{ij}
+\psi^{,k}_{\ ,k}\gamma_{ij}.\nonumber
\end{eqnarray}

The Ricci scalar $R$ and relevant components of Einstein tensor
$G_{ab}=R_{ab}-\frac{1}{2}g_{ab}R$ are
\begin{eqnarray}
R &=&
\frac{6}{a^2}\left[\psi''-\frac{a''}{a}(1-2\phi)+\frac{a'}{a}\left(\phi'+3\psi'\right)\right]\nonumber\\
&&-\frac{1}{a^2}\left[4\psi^{,i}_{\ ,i}-2\phi^{,i}_{\
,i}\right],\\
G^{0}_{\ 0} &=& \frac{3}{a^2}\left(\frac{a'}{a}\right)^2(1-2\phi)
-\frac{6}{a^2}\frac{a'}{a}\psi' + \frac{2}{a^2}\psi^{,i}_{\ ,i},\\
G^{i}_{\ j} &=&
\frac{1}{a^2}\left[-2\psi''-2\frac{a'}{a}\left(\phi'+2\psi'\right)
-(\phi-\psi)^{,k}_{\ ,k}\right]\delta^{i}_{\ j}\nonumber\\
&&+\frac{1}{a^2}\left[2\frac{a''}{a}-\left(\frac{a'}{a}\right)^2\right](1-2\phi)\delta^{i}_{\
j}\nonumber\\
&&+\frac{1}{a^2}\left(\phi-\psi\right)^{,i}_{\ ,j}.
\end{eqnarray}

\section{Discrete Equations for the $N$-body Simulations}
\label{appen:discret}

In the \texttt{MLAPM} code the Poisson equation Eq.~(\ref{eq:WFPoisson}) is
(and in our modified code the scalar field equation of motion Eq.~(\ref{eq:WFphiEOM}) will also be) solved on discretised grid points, so we must
develop the discrete versions of Eqs.~(\ref{eq:WFphiEOM}, \ref{eq:WFPoisson}) to be implemented in the code. Before doing that, we note that
Eqs.~(\ref{eq:WFphiEOM}, \ref{eq:WFPoisson}) are not independent but
are coupled together, which could further complicate the solver. As a
result, we should first decouple them by eliminating $\vec{\partial}_{\mathbf{x}}^{2}\left( a\sqrt{\kappa_{\ast}}\delta\varphi\right)$ 
($\vec{\partial}_{\mathbf{x}}^{2}\Phi$) from the equation for $\Phi$ ($\delta
\varphi $). This is easy to do and the resulted equations are respectively  
\begin{widetext}
\begin{eqnarray}\label{eq:WFphiEOM2}
\left[1+\frac{6\gamma^2\kappa_{\ast}\bar{\varphi}^2}
{1+\gamma\kappa_{\ast}\bar{\varphi}^2}\right]
c^2\vec{\partial}^{2}_{\mathbf{x}}\left(a\sqrt{\kappa_{\ast}}\delta\varphi\right)
&=&
-6\gamma\left(\mathcal{H}'+\mathcal{H}^2\right)a\sqrt{\kappa_{\ast}}\delta\varphi
-3\alpha\lambda
H_0^2a^{3}\left[\frac{1}{\left(\sqrt{\kappa_{\ast}}\varphi\right)^{\alpha}}
-\frac{1}{\left(\sqrt{\kappa_{\ast}}\bar{\varphi}\right)^{\alpha}}\right]\nonumber\\
&&-3\gamma\sqrt{\kappa_{\ast}}\bar{\varphi}\left(1+\gamma\kappa_{\ast}\bar{\varphi}^2_0\right)\Omega_mH_0^2
\left[\frac{\rho_c}{1+\gamma\kappa_{\ast}\varphi^2}
-\frac{1}{1+\gamma\kappa_{\ast}\bar{\varphi}^2}\right]\nonumber\\
&&+6\gamma\sqrt{\kappa_{\ast}}\bar{\varphi}\lambda H_0^2a^3
\left[\frac{1}{\left(1+\gamma\kappa_{\ast}\varphi^2\right)\left(\sqrt{\kappa_{\ast}}\varphi\right)^\alpha}-
\frac{1}{\left(1+\gamma\kappa_{\ast}\bar{\varphi}^2\right)\left(\sqrt{\kappa_{\ast}}\bar{\varphi}\right)^\alpha}\right]\nonumber\\
&&-2\gamma\sqrt{\kappa_{\ast}}\bar{\varphi}a\left[(1+3\gamma)\kappa_{\ast}\bar{\varphi}'^2
+3\gamma\kappa_{\ast}\bar{\varphi}\bar{\varphi}''\right]
\left[\frac{1}{1+\gamma\kappa_{\ast}\varphi^2}-\frac{1}{1+\gamma\kappa_{\ast}\bar{\varphi}^2}\right]
\end{eqnarray}
\end{widetext}
for the scalar field, and
\begin{widetext}
\begin{eqnarray}\label{eq:WFPoisson2}
&&\frac{1+\gamma\kappa_{\ast}\bar{\varphi}^2+6\gamma^2\kappa_{\ast}\bar{\varphi}^2}
{1+\gamma\kappa_{\ast}\bar{\varphi}^2+8\gamma^2\kappa_{\ast}\bar{\varphi}^2}
\vec{\partial}^{2}_{\mathbf{x}}\Phi\nonumber\\
&=&\frac{3}{2}\left(1+\gamma\kappa_{\ast}\bar{\varphi}^2_0\right)\Omega_mH_0^2
\left[\frac{\rho_c}{1+\gamma\kappa_{\ast}\varphi^2}
-\frac{1}{1+\gamma\kappa_{\ast}\bar{\varphi}^2}\right]-3\lambda
H_0^2a^3
\left[\frac{1}{\left(1+\gamma\kappa_{\ast}\varphi^2\right)\left(\sqrt{\kappa_{\ast}}\varphi\right)^\alpha}-
\frac{1}{\left(1+\gamma\kappa_{\ast}\bar{\varphi}^2\right)\left(\sqrt{\kappa_{\ast}}\bar{\varphi}\right)^\alpha}\right]
\nonumber\\
&& +a\left[(1+3\gamma)\kappa_{\ast}\bar{\varphi}'^2
+3\gamma\kappa_{\ast}\bar{\varphi}\bar{\varphi}''\right]
\left[\frac{1}{1+\gamma\kappa_{\ast}\varphi^2}-\frac{1}{1+\gamma\kappa_{\ast}\bar{\varphi}^2}\right]
+\frac{6\gamma^2\sqrt{\kappa_{\ast}}\bar{\varphi}}
{1+\gamma\kappa_{\ast}\bar{\varphi}^2+8\gamma^2\kappa_{\ast}\bar{\varphi}^2}
\left(\mathcal{H}'+\mathcal{H}^2\right)a\sqrt{\kappa_\ast}\delta\varphi\nonumber\\
&&+\frac{3\gamma\alpha\lambda\sqrt{\kappa_{\ast}}\bar{\varphi}}
{1+\gamma\kappa_{\ast}\bar{\varphi}^2+8\gamma^2\kappa_{\ast}\bar{\varphi}^2}H_0^2a^3
\left[\frac{1}{\left(\sqrt{\kappa_{\ast}}\varphi\right)^{1+\alpha}}
-\frac{1}{\left(\sqrt{\kappa_{\ast}}\bar{\varphi}\right)^{1+\alpha}}\right]
\end{eqnarray}
\end{widetext}
for the gravitational potential.

Introducing the variable $u$ (cf.~Sect.~\ref{subsect:codeunit}),
the Poisson equation becomes
\begin{widetext}
\begin{eqnarray}\label{eq:u_Poisson}
&&\frac{1+\gamma\kappa_{\ast}\bar{\varphi}^2+6\gamma^2\kappa_{\ast}\bar{\varphi}^2}
{1+\gamma\kappa_{\ast}\bar{\varphi}^2+8\gamma^2\kappa_{\ast}\bar{\varphi}^2}
\nabla^{2}\Phi_{c}\\
&=&
\frac{3}{2}\left(1+\gamma\kappa_\ast\bar{\varphi}^{2}_{0}\right)\Omega_m
\left[\frac{\rho_c}{1+\gamma\left(\sqrt{\kappa_{\ast}}\bar{\varphi}+\frac{B^2H_0^2}{ac^2}u\right)^2}
-\frac{1}{1+\gamma\kappa_{\ast}\bar{\varphi}^2}\right]\nonumber\\
&&-3\lambda
a^3\left[\frac{1}{1+\gamma\left(\sqrt{\kappa_{\ast}}\bar{\varphi}+\frac{B^2H_0^2}{ac^2}u\right)^2}
\frac{1}{\left(\sqrt{\kappa_{\ast}}\bar{\varphi}+\frac{B^2H_0^2}{ac^2}u\right)^\alpha}
-\frac{1}{1+\gamma\kappa_{\ast}\bar{\varphi}^2}
\frac{1}{\left(\sqrt{\kappa_{\ast}}\bar{\varphi}\right)^\alpha}\right]\nonumber\\
&&+a\left[(1+3\gamma)\frac{\kappa_{\ast}\bar{\varphi}'^2}{H_0^2}+
3\gamma\sqrt{\kappa}_{\ast}\bar{\varphi}\frac{\sqrt{\kappa_\ast}\bar{\varphi}''}{H_0^2}\right]
\left[\frac{1}{1+\gamma\left(\sqrt{\kappa_{\ast}}\bar{\varphi}+\frac{B^2H_0^2}{ac^2}u\right)^2}
-\frac{1}{1+\gamma\kappa_{\ast}\bar{\varphi}^2}\right]\nonumber\\
&&+\frac{\gamma\sqrt{\kappa_{\ast}}\bar{\varphi}}{1+\gamma\kappa_\ast\varphi^2+8\gamma^2\kappa_\ast\varphi^2}
\left\{6\gamma\left[\frac{\mathcal{H}'}{H_0^2}+\frac{\mathcal{H}^2}{H_0^2}\right]
\frac{\left(BH_0\right)^2}{c^2}u + 3\alpha\lambda
a^3\left[\frac{1}{\left(\sqrt{\kappa_{\ast}}\bar{\varphi}+\frac{B^2H_0^2}{ac^2}u\right)^{1+\alpha}}
-\frac{1}{\left(\sqrt{\kappa_{\ast}}\bar{\varphi}\right)^{1+\alpha}}\right]\right\}
\end{eqnarray}
\end{widetext}
where $\lambda$ is defined in Sect.~\ref{subsect:codeunit} and is
a constant of $\mathcal{O}(1)$. We have also used the code unit
for other quantities. This equation contains $u$, which must be
solved from the scalar field equation of motion.

The scalar field equation of motion can be similarly written. 
In order that the equation can be integrated into \texttt{MLAPM}, we need
to discretise it for the application of Newton-Gauss-Seidel relaxation
method. This means writing down a discrete version of this equation
on a uniform grid with grid spacing $h$. Suppose we want to achieve
second-order precision, as is in the default Poisson solver of \texttt{MLAPM}, then $\nabla^{2}u$ in one dimension can be written as
\begin{eqnarray}
\nabla^{2}u &\rightarrow& \nabla^{h2}u_{j}\ =\
\frac{u_{j+1}+u_{j-1}-2u_{j}}{h^2}
\end{eqnarray}
where a subscript $_{j}$ means that the quantity is evaluated on
the $j$-th point. The generalisation to three dimensions is
straightforward.

The discrete version of the equation of motion for $u$ is then
\begin{eqnarray}\label{eq:diffop}
L^{h}\left(u_{i,j,k}\right) &=& 0,
\end{eqnarray}
in which
\begin{widetext}
\begin{eqnarray}\label{eq:u_phi_EOM}
L^{h}\left(u_{i,j,k}\right) &=&
\frac{1+\gamma\kappa_{\ast}\bar{\varphi}^2+6\gamma^2\kappa_{\ast}\bar{\varphi}^2}
{1+\gamma\kappa_{\ast}\bar{\varphi}^2}\frac{1}{h^{2}}
\left[u_{i+1,j,k}+u_{i-1,j,k}+u_{i,j+1,k}+u_{i,j-1,k}+u_{i,j,k+1}+u_{i,j,k-1}-6u_{i,j,k}\right]\nonumber\\
&&+3\gamma\sqrt{\kappa_\ast}\bar{\varphi}\left(1+\gamma\kappa_\ast\bar{\varphi}^{2}_0\right)
\Omega_m\left[\frac{\rho_c}{1+\gamma\left(\sqrt{\kappa_{\ast}}\bar{\varphi}+\frac{B^2H_0^2}{ac^2}u_{i,j,k}\right)^2}
-\frac{1}{1+\gamma\kappa_{\ast}\bar{\varphi}^2}\right]\nonumber\\
&&+3\alpha\lambda a^3
\left[\frac{1}{\left(\sqrt{\kappa_{\ast}}\bar{\varphi}+\frac{B^2H_0^2}{ac^2}u_{i,j,k}\right)^{1+\alpha}}
-\frac{1}{\left(\sqrt{\kappa_{\ast}}\bar{\varphi}\right)^{1+\alpha}}\right]
+6\gamma\frac{\mathcal{H'}+\mathcal{H}^2}{H_0^2}\frac{B^2H_0^2}{c^2}u_{i,j,k}\nonumber\\
&&-6\gamma\sqrt{\kappa_\ast}\bar{\varphi}\lambda a^3
\left[\frac{1}{1+\gamma\left(\sqrt{\kappa_{\ast}}\bar{\varphi}+\frac{B^2H_0^2}{ac^2}u_{i,j,k}\right)^2}
\frac{1}{\left(\sqrt{\kappa_{\ast}}\bar{\varphi}+\frac{B^2H_0^2}{ac^2}u_{i,j,k}\right)^\alpha}
-\frac{1}{1+\gamma\kappa_{\ast}\bar{\varphi}^2}
\frac{1}{\left(\sqrt{\kappa_{\ast}}\bar{\varphi}\right)^\alpha}\right]\nonumber\\
&&+2\gamma\sqrt{\kappa_\ast}\bar{\varphi}a\left[(1+3\gamma)\frac{\kappa_{\ast}\bar{\varphi}'^2}{H_0^2}+
3\gamma\sqrt{\kappa}_{\ast}\bar{\varphi}\frac{\sqrt{\kappa_\ast}\bar{\varphi}''}{H_0^2}\right]
\left[\frac{1}{1+\gamma\left(\sqrt{\kappa_{\ast}}\bar{\varphi}+\frac{B^2H_0^2}{ac^2}u_{i,j,k}\right)^2}
-\frac{1}{1+\gamma\kappa_{\ast}\bar{\varphi}^2}\right]
\end{eqnarray}
\end{widetext}
Then, the Newton-Gauss-Seidel iteration says that we can
obtain a new (and often more accurate) solution of $u$, $u_{i,j,k}^{\mathrm{%
new}}$, using our knowledge about the old (and less accurate) solution $%
u_{i,j,k}^{\mathrm{old}}$ via
\begin{eqnarray}\label{eq:GS}
u^{\mathrm{new}}_{i,j,k} &=& u^{\mathrm{old}}_{i,j,k} -
\frac{L^{h}\left(u^{\mathrm{old}}_{i,j,k}\right)}{\partial
L^{h}\left(u^{\mathrm{old}}_{i,j,k}\right)/\partial u_{i,j,k}}.
\end{eqnarray}
The old solution will be replaced by the new solution to
$u_{i,j,k}$ once the new solution is ready, using the red-black
Gauss-Seidel sweeping scheme. Note that
\begin{widetext}
\begin{eqnarray}
\frac{\partial L^{h}(u_{i,j,k})}{\partial u_{i,j,k}} &=&
-\frac{1+\gamma\kappa_{\ast}\bar{\varphi}^2+6\gamma^2\kappa_{\ast}\bar{\varphi}^2}
{1+\gamma\kappa_{\ast}\bar{\varphi}^2}\frac{6}{h^{2}}
+6\gamma\frac{\mathcal{H'}+\mathcal{H}^2}{H_0^2}\frac{B^2H_0^2}{c^2}
-\frac{3\alpha(1+\alpha)\lambda
a^2\left(BH_0/c\right)^2}{\left(\sqrt{\kappa_{\ast}}\bar{\varphi}+\frac{B^2H_0^2}{ac^2}u_{i,j,k}\right)^{2+\alpha}}\nonumber\\
&&-6\gamma^2\frac{B^2H_0^2}{ac^2}\sqrt{\kappa_\ast}\bar{\varphi}\left(1+\gamma\kappa_\ast\bar{\varphi}^{2}_0\right)\Omega_m\rho_c
\frac{\sqrt{\kappa_{\ast}}\bar{\varphi}+\frac{B^2H_0^2}{ac^2}u_{i,j,k}}
{\left[1+\gamma\left(\sqrt{\kappa_{\ast}}\bar{\varphi}+\frac{B^2H_0^2}{ac^2}u_{i,j,k}\right)^2\right]^2}\nonumber\\
&&+12\gamma^2\frac{B^2H_0^2}{ac^2}\lambda\sqrt{\kappa_{\ast}}\bar{\varphi}
a^3\frac{\sqrt{\kappa_{\ast}}\bar{\varphi}+\frac{B^2H_0^2}{ac^2}u_{i,j,k}}
{\left[1+\gamma\left(\sqrt{\kappa_{\ast}}\bar{\varphi}+\frac{B^2H_0^2}{ac^2}u_{i,j,k}\right)^2\right]^2}
\frac{1}{\left(\sqrt{\kappa_{\ast}}\bar{\varphi}+\frac{B^2H_0^2}{ac^2}u_{i,j,k}\right)^\alpha}\nonumber\\
&&+6\alpha\gamma\frac{B^2H_0^2}{ac^2}\lambda\sqrt{\kappa_{\ast}}\bar{\varphi}
a^3\frac{1}{1+\gamma\left(\sqrt{\kappa_{\ast}}\bar{\varphi}+\frac{B^2H_0^2}{ac^2}u_{i,j,k}\right)^2}
\frac{1}{\left(\sqrt{\kappa_{\ast}}\bar{\varphi}+\frac{B^2H_0^2}{ac^2}u_{i,j,k}\right)^{1+\alpha}}\nonumber\\
&&-4\gamma^2\frac{B^2H_0^2}{c^2}\sqrt{\kappa_{\ast}}\bar{\varphi}
\left[(1+3\gamma)\frac{\kappa_{\ast}\bar{\varphi}'^2}{H_0^2}+
3\gamma\sqrt{\kappa}_{\ast}\bar{\varphi}\frac{\sqrt{\kappa_\ast}\bar{\varphi}''}{H_0^2}\right]
\frac{\sqrt{\kappa_{\ast}}\bar{\varphi}+\frac{B^2H_0^2}{ac^2}u_{i,j,k}}
{\left[1+\gamma\left(\sqrt{\kappa_{\ast}}\bar{\varphi}+\frac{B^2H_0^2}{ac^2}u_{i,j,k}\right)^2\right]^2}.
\end{eqnarray}
\end{widetext}

In principle, if we start from a high redshift, then the initial guess of $u_{i,j,k}$ for the relaxation can be so chosen that the initial value of 
$\varphi $ in all space is equal to the background value
$\bar{\varphi}$, because at this time we expect this to be
approximately true any way. At subsequent time-steps we could use
the solution for $u_{i,j,k}$ from the previous time-step as our
initial guess. If the timestep is small enough then we would
expect $u$ to change only slightly between consecutive timesteps
so that such a guess will be good enough for the iterations to
converge quickly.

\section{Algorithm to Solve the Background Evolution}

\label{appen:bkgd}

Here we give our formulae and algorithm for the background field
equations which can also be applied to linear Boltzmann codes such
as \texttt{CAMB}. Throughout this Appendix we use the conformal
time $\tau$ instead of the physical time $t$, and $' \equiv
d/d\tau ,\mathcal{H}\equiv a'/a$. All quantities appearing here
are background ones unless stated otherwise.

For convenience, we will work with dimensionless quantities and
define $\psi\equiv \sqrt{\kappa_{\ast}}\varphi $ and $N\equiv \ln
a$ so that
\begin{eqnarray}
\psi' &=& \mathcal{H}\frac{d\psi}{dN},\\
\psi'' &=& \mathcal{H}^{2}\frac{d^2\psi}{dN^2} +
\mathcal{H}'\frac{d\psi}{dN}.
\end{eqnarray}
With these definitions it is straightforward to show that the
scalar field equation of motion can be expressed as
\begin{eqnarray}\label{eq:bkgd_scalar2}
\left(\frac{\mathcal{H}}{\mathcal{H}_{0}}\right)^2\frac{d^{2}\psi}{dN^{2}}
+
\left(2\frac{\mathcal{H}^2}{\mathcal{H}_0^{2}}+\frac{\mathcal{H}'}{\mathcal{H}_0^{2}}\right)\frac{d\psi}{dN}\nonumber\\
+ \frac{\kappa_{\ast}}{\mathcal{H}^2_0}a^{2}\frac{\partial
V(\psi)}{\partial\psi} -
3\left(\frac{\mathcal{H}'}{\mathcal{H}_0^2}+\frac{\mathcal{H}^2}{\mathcal{H}_0^2}\right)\frac{\partial
f(\psi)}{\partial\psi} &=& 0,
\end{eqnarray}
where $\mathcal{H}_{0}$ is the current value of $\mathcal{H}$.

Obviously we need to know how to compute the quantities
$\mathcal{H}/\mathcal{H}_0$ and $\mathcal{H}'/\mathcal{H}_0$ as
well. For $\mathcal{H}/\mathcal{H}_0$, we start with the Friedmann
equation
\begin{eqnarray}\label{eq:bkgd_friedman}
3\mathcal{H}^2 &=&
\frac{1}{1+f}\kappa_{\ast}\left[\rho_{m}+\rho_{r}+V(\psi)\right]a^{2}\nonumber\\
&&
+\frac{1}{1+f}\left[\frac{1}{2}\left(\frac{d\psi}{dN}\right)^2-3\frac{df}{d\psi}\frac{d\psi}{dN}\right]\mathcal{H}^{2},
\end{eqnarray}
where $\rho_m$ and $\rho_r$ are the energy densities for matter and
radiation respectively. We define the fractional energy densities
for matter and radiation respectively as
\begin{eqnarray}
\label{eq:bkgd_omgm}\Omega_{m} &\equiv&
\frac{\kappa_{eff}\rho_{m0}}{3\mathcal{H}^{2}_{0}}\ =\
\frac{1}{1+f_0}\frac{\kappa_{\ast}\rho_{m0}}{3\mathcal{H}^{2}_{0}},\\
\label{eq:bkgd_omgr}\Omega_{r} &\equiv&
\frac{\kappa_{\bigoplus0}\rho_{r0}}{3\mathcal{H}^{2}_{0}}\nonumber\\
&=&
\frac{1}{1+f_0}\frac{2\left(1+f_0\right)+4\left(\frac{df}{d\psi}\right)_0^2}
{2\left(1+f_0\right)+3\left(\frac{df}{d\psi}\right)_0^2}\frac{\kappa_{\ast}\rho_{r0}}{3\mathcal{H}^{2}_{0}},
\end{eqnarray}
where a subscript $_0$ means the present-day value. Notice the
difference between these definitions, which comes from the different
treatments for radiation and matter in numerical codes such as
{\tt CAMB}. For radiation, \emph{e.g.}, photon, we know the
present temperature of the CMB and thus its exact energy density
$\rho_{r0}$, as well as the locally measured value of
gravitational constant $\kappa_{\bigoplus0}$ (which in scalar-tensor theories is in general different from $\kappa_{\ast}$) and
current Hubble expansion rate $\mathcal{H}_0$, and so the
definition Eq.~(\ref{eq:bkgd_omgr}) comes out naturally, where we
have used the relation between $\kappa_{\ast}$ and
$\kappa_{\bigoplus}$ \footnotemark[1]. For matter, the
fractional energy density is to be interpreted from the
cosmological observables such as CMB and large scale structure,
which are obviously different in $\Lambda$CDM and scalar-tensor
theories; consequently there is some freedom in defining it and we
make it as in Eq.~(\ref{eq:bkgd_omgm}).

\footnotetext[1]{(Massless) neutrinos are treated similarly, but
the neutrino background has a temperature lower than that of CMB,
due to the energy transfer into photons during the
electron-positron annihilation, but not into neutrinos which have
decoupled by then.}

Then, remembering that
\begin{eqnarray}
\rho_{m} &\propto& a^{-3},\\
\rho_r &\propto& a^{-4},
\end{eqnarray}
we have
\begin{eqnarray}\label{eq:bkgd_friedman2}
\left(\frac{\mathcal{H}}{\mathcal{H}_0}\right)^2 &=&
\frac{\frac{\kappa_{\ast}}{\kappa_{\bigoplus0}}\Omega_{r}a^{-2}+\left(1+f_0\right)\Omega_{m}a^{-1}+\frac{\kappa_{\ast}Va^{2}}{3\mathcal{H}^2_0}}
{1+f+\frac{df}{d\psi}\frac{d\psi}{dN}-\frac{1}{6}\left(\frac{d\psi}{dN}\right)^2},\
\ \ \
\end{eqnarray}
in which (where both $\kappa_{\ast}$ and $\kappa_{\bigoplus0}$
are constants, and $\kappa_{\bigoplus0}$ is the present value of
$\kappa_{\bigoplus}$)
\begin{eqnarray}
\frac{\kappa_{\ast}}{\kappa_{\bigoplus0}} &=& \left(1+f_0\right)
\frac{2\left(1+f_0\right)+3\left(\frac{df}{d\psi}\right)_0^2}
{2\left(1+f_0\right)+4\left(\frac{df}{d\psi}\right)_0^2}.
\end{eqnarray}
For $\mathcal{H}'/\mathcal{H}_{0}$, we use the Raychaudhrui
equation
\begin{eqnarray}
\mathcal{H}' &=&
-\frac{1}{6}\kappa_{\ast}\left(\rho+3p\right)a^2\nonumber\\
&=&
-\frac{1}{6}\frac{1}{1+f}\kappa_\ast\left[\rho_m+2\rho_r-2V(\psi)\right]a^2\nonumber\\
&&-\frac{1}{6}\frac{1}{1+f}\left(2+3\frac{d^2f}{d\psi^2}\right)\left(\frac{d\psi}{dN}\right)^2\mathcal{H}^2\nonumber\\
&&-\frac{1}{2}\frac{1}{1+f}\frac{df}{d\psi}\left(\frac{d\psi}{dN}\mathcal{H}'+\frac{d^{2}\psi}{dN^2}\mathcal{H}^{2}\right).
\end{eqnarray}
As in the above, dividing this by $\mathcal{H}^2_0$ and
rearranging, we obtain
\begin{eqnarray}\label{eq:bkgd_raychaudhuri2}
\frac{\mathcal{H}'}{\mathcal{H}^{2}_0} &=&
-\frac{\frac{1}{2}\left[\left(1+f_0\right)\Omega_ma^{-1}+2\frac{\kappa_{\ast}}{\kappa_{\bigoplus0}}\Omega_ra^{-2}\right]-\frac{\kappa_{\ast}Va^2}{3\mathcal{H}^2_0}}
{1+f+\frac{1}{2}\frac{df}{d\psi}\frac{d\psi}{dN}}\nonumber\\
&&-\frac{\frac{1}{2}\frac{df}{d\psi}\frac{d^{2}\psi}{dN^2}
+\left(\frac{1}{3}+\frac{1}{2}\frac{d^2f}{d\psi^2}\right)\left(\frac{d\psi}{dN}\right)^2}
{1+f+\frac{1}{2}\frac{df}{d\psi}\frac{d\psi}{dN}}\frac{\mathcal{H}^2}{\mathcal{H}_0^2}.
\end{eqnarray}

Substituting Eqs.~(\ref{eq:bkgd_friedman2},
\ref{eq:bkgd_raychaudhuri2}) into Eq.~(\ref{eq:bkgd_scalar2}), we
finally arrive at
\begin{eqnarray}
\frac{1+f+\frac{3}{2}\left(\frac{df}{d\psi}\right)^2}{1+f+\frac{1}{2}\frac{df}{d\psi}\frac{d\psi}{dN}}A\frac{d^2\psi}{dN^2}
+ \left(2A+B\right)\frac{d\psi}{dN}\nonumber\\
+ \frac{\kappa_{\ast}}{\mathcal{H}_0^2}\frac{dV}{d\psi}a^2 -
3(A+B)\frac{df}{d\psi} &=& 0, \ \
\end{eqnarray}
in which we have defined
\begin{eqnarray}
A &\equiv& \frac{\mathcal{H}^2}{\mathcal{H}^2_0},\\
B &\equiv& \frac{\mathcal{H}'}{\mathcal{H}^2_0} +
\frac{\frac{1}{2}\frac{df}{d\psi}\frac{\mathcal{H}^2}{\mathcal{H}^2_0}}{1+f+\frac{1}{2}\frac{df}{d\psi}\frac{d\psi}{dN}}\frac{d^{2}\psi}{dN^2},
\end{eqnarray}
where $A$ and $B$ do not contain $d^2\psi/dN^2$, to lighten the
notation.

When solving for $\varphi $ (or $\psi$), we just use
Eq.~(\ref{eq:bkgd_scalar2}) aided by
Eqs.~(\ref{eq:bkgd_friedman2}, \ref{eq:bkgd_raychaudhuri2}). It
may appear then that, given any initial values for
$\psi_{\mathrm{ini}}$ and $\left( d\psi/dN\right)_{\mathrm{ini}},$
the evolution of $\varphi$ is obtainable. However,
Eq.~(\ref{eq:bkgd_friedman2}) is not necessarily satisfied for
$\psi$ evolved in such way. Instead, it constrains the initial
condition $\psi$ must start with, and the way it must subsequently
evolve. This in turn is determined by the parameters $\lambda,
\alpha, \gamma$, since $\alpha, \gamma$ specify a model and are
fixed once the model is chosen; the only concern is $\lambda$.

As for the initial conditions $\psi_{\mathrm{ini}}$ and $\left( d%
\psi/dN\right) _{\mathrm{ini}}$, we have found that the subsequent
evolution of $\psi$ is rather insensitive to them.
Thus, we choose $\psi_{\mathrm{ini}}=\left( d\psi%
/dN\right) _{\mathrm{ini}}=0$ at some very early time (say
$N_{\mathrm{ini}}$ corresponds to
$a_{\mathrm{ini}}=e^{N_{\mathrm{ini}}}=10^{-8}$) in all the
models. Such a choice is clearly not only practical but also
reasonable, given the fact that we expect that the scalar field
starts high up the potential and rolls down subsequently.

As for $\lambda $, we use a trial-and-error method to find its
value which ensures that (again subscript $_0$ indicates the
current time)
\begin{eqnarray}
\frac{\frac{\kappa_{\ast}}{\kappa_{\bigoplus0}}\Omega_{r}+\left(1+f_0\right)\Omega_{m}+\frac{\kappa_{\ast}V_0}{3\mathcal{H}^2_0}}
{1+f_0+\left(\frac{df}{d\psi}\right)_0\left(\frac{d\psi}{dN}\right)_0-\frac{1}{6}\left(\frac{d\psi}{dN}\right)_0^2}
&=& 1
\end{eqnarray}
which comes from setting $a=1$ in Eq.~(\ref{eq:bkgd_friedman2}).

We determine the correct value of $\lambda $ for any given $\alpha
, \gamma$ in this way using a trial-and-error routine, and then
compute the values of $\psi$ and $d\psi/dN$ for predefined values
of $N$ stored in an array. Their values at any time are then
obtained using interpolation, and with these it is straightforward
to compute other relevant quantities, such as $\mathcal{H},
\mathcal{H}'$ and $\varphi$, which are used in the linear
perturbation computations.

\bibliography{alpha,nbody}

\end{document}